\documentclass[a4paper, 11pt]{article}
\usepackage[hmargin=2.5cm,vmargin=3.5cm]{geometry}
\usepackage[utf8]{inputenc}
\usepackage{amsmath}
\usepackage{caption}   %% HUGE PACKAGE FOR TABLE  CAPTION!!
\usepackage{mathrsfs}

\usepackage{graphicx}
\usepackage{subfig}
\usepackage{hyperref}
\usepackage{multicol}
\usepackage{mathtools}
\usepackage{eurosym}
\usepackage{pdfpages}
\usepackage[normalem]{ulem}
\usepackage{latexsym}
\usepackage{amssymb}
\usepackage[capposition=top]{floatrow}
\usepackage{lipsum}
\usepackage{setspace}
\usepackage{float}
\usepackage[english]{babel}
\usepackage{fancyhdr, afterpage}
\usepackage{graphicx}
\usepackage{latexsym}
\usepackage{bbm}
\usepackage{amsmath}
\usepackage{amsfonts}
\usepackage{amssymb}
\usepackage[utf8]{inputenc}
\usepackage{caption}
\usepackage{bm}
\usepackage{pdflscape} 
\usepackage{subfig}
\usepackage{tikz}
\usepackage{csquotes}
\usepackage[subnum]{cases}
\usepackage{mathtools}% Loads amsmath
\DeclareMathOperator*{\E}{\mathbb{E}}
\usepackage{booktabs}
\hypersetup{
colorlinks=true,
linkcolor=blue,
filecolor=black, 
urlcolor=blue,
citecolor=blue
}
\usepackage[authordate,bibencoding =auto,strict,backend=biber,natbib, maxcitenames=1,uniquename=false,
  uniquelist=false, doi=false,isbn=false,url=false,eprint=false]{biblatex-chicago}
\renewcommand{\footnoterule}{%
  \kern -3pt
  \hrule width \textwidth height 1pt
  \kern 2pt
}
\title{Modelling and Forecasting Macroeconomic Risk with Time Varying Skewness Stochastic Volatility Models
\thanks{This paper is part of my PhD dissertation at the University of Bologna. I am grateful to Andrea Carriero for invaluable support and guidance. I thank Luca Fanelli and Umberto Cherubini for comments and suggestions and participants at the March 2021 Unibo PhD Forum, the July  2022 Unibo PhD Forum and the Oslo IAAE 2023 conference.}}
\author{Andrea Renzetti\footnote{Department of Economics, Alma Mater Studiorium Università di Bologna, Piazza Scaravilli 2, 40126 Bologna, Italy}}
\date{ \small First Draft: September 21, 2021 \\ 
This Draft: \today \\
\href{https://drive.google.com/file/d/17O3IkaK7TF-L8k2Mv9u8SuaPKUbt-Q7f/view?usp=drive_link}{Latest Draft Here}}
\begin{document}
\maketitle
\begin{abstract}
Monitoring downside risk and upside risk to the key macroeconomic indicators is critical for effective policymaking aimed at maintaining economic stability. In this paper I propose a parametric framework for modelling and forecasting macroeconomic risk based on stochastic volatility models with \textit{Skew-Normal} and \textit{Skew-t} shocks featuring time varying skewness. Exploiting a mixture stochastic representation of the \textit{Skew-Normal} and \textit{Skew-t}  random variables, in the paper I develop efficient posterior simulation samplers for Bayesian estimation of both univariate and VAR models of this type. In an application, I use the models to predict downside risk to GDP growth in the US and I show that these models represent a competitive alternative to semi-parametric approaches such as quantile regression. Finally, estimating a medium scale VAR on US data I show that time varying skewness is a relevant feature of macroeconomic and financial shocks. \\

\emph{J.E.L Classification Code:  C22, C32, C53 }

\bigskip
\bigskip
\bigskip
\bigskip
\end{abstract}
\clearpage
%\section*{Non Technical Summary}

%Policymakers have been increasingly focusing on monitoring and forecasting tail risks in macroeconomic outcomes. The interest in tail risks reflects an underlying perception or assumption of asymmetries in distributions of outcomes.This article contributes to the recently growing body of research that focuses on predicting and assessing tail risk in macroeconomic outcomes by showing how to directly model changes in conditional skewness of macroeconomic variables. It extends the Univariate Stochastic Volatility Model \citep{jpr1993} and the Vector Autoregressive Model (VAR) with Stochastic Volatility \citep{COGLEY2005262} to model asymmetries in the predictive distribution of the outcome variables. In contrast to the vast growing body of studies that used quantile regression methods to assess and predict macroeconomic risk, this approach for modelling and forecasting tail risk in the outcome of multiple macroeconomic time series is fully parametric and allows to take advantage of all the familiar toolkits used for structural analysis that came with VARs. 

\clearpage

\section{Introduction}

Central banks and policy institutions play a critical role in maintaining financial stability and fostering economic growth. A key challenge they face is effectively monitoring the likelihood of severe events that could have adverse effects on the economy. Failing to adequately assess these risks can lead to underestimation of potential losses and insufficient policy responses. To address this challenge, it is essential to develop econometric tools that can accurately predict and assess tail risk in macroeconomic outcomes. In this paper, I propose an econometric framework specifically designed for modeling and forecasting macroeconomic tail risk. The framework relies on fully parametric univariate and multivariate stochastic volatility models with \textit{Skew-Normal} and \textit{Skew-t} shocks  featuring stochastic skewness. These models aim to capture and predict persistent time-varying asymmetries in the future distribution of the variables of interest. Capturing these asymmetries is especially relevant given the risk management nature  of the problem of policymaking faced by central banks and policy institutions \citep{km}.

The paper begins by extending the well-known univariate stochastic volatility model introduced by \citet{jpr1993} to explicitly account for time-varying conditional skewness in the predictive distribution of a single target variable. Then, building upon the univariate approach, the paper introduces a Bayesian Vector Autoregressive (VAR) model with stochastic volatility and time-varying skewness.  By allowing to track changes in the shape of the predictive distribution of multiple time series, this model is suitable for quantification and forecasting of tail risk to multiple target variables. Importantly, the model retains all the advantages and familiar toolkit for policy analysis and scenario analysis associated to the VAR framework.  The model is estimated through an efficient \textit{Gibbs sampler} that exploits a convenient mixture stochastic representation of the \textit{Skew-Normal} and \textit{Skew-t} shocks. To test the effectiveness of the proposed framework, I use the time-varying skewness stochastic volatility models to monitor downside risk to GDP growth in the US economy. The findings of this analysis align with the main conclusions of \citet{ABG}, revealing a nonlinear and asymmetric impact of financial conditions on the future distribution of GDP growth. Additionally, the models provide slightly more accurate out-of-sample forecasts of downside risk compared to quantile regression, which is often considered as the benchmark model in this literature. Furthermore, estimating a medium-scale VAR model of monetary policy, I show that shocks to financial and macroeconomic time series exhibit both time-varying volatility and time-varying skewness, suggesting that taking into consideration both of these features might be of particularly relevance for accurately assessing upside and downside risk to macroeconomic indicators. 

\textbf{Related literature} \hspace{0.2cm} A fast-growing body of studies recently used univariate quantile regression methods for modelling and predicting asymmetries in the future distribution of the macroeconomic variables of interest. For example, \citet{GIGLIO2016457} used predictive quantile regression to investigate whether systemic risk indicator and financial distress indicators predict changes in the lower quantiles of future macroeconomics shocks. As well, \citet{kiley2018unemployment} used quantile regression to examine fluctuations in the risk of a large increase in unemployment. More recently, \citet{ABG} used a two step-procedure based on predictive quantile regression and quantile interpolation to model changes in downside risk to future GDP growth as a function of current financial and economic conditions.\footnote{The two step approach based on quantile regression of \citet{ABG} gained substantial popularity in the literature and has been employed in many other frameworks to assess and predict tail risk to economic outcomes. Among the others, \citet{loria2019} used the two step approach of \citet{ABG} for assessing and predicting downside and upside risk to inflation while \citet{GELOS2022103555} used the same approach for predicting the probability of large capital out-flows and in-flows to emerging markets.} Despite its popularity, the quantile regression method of \citet{ABG} typically fails in the presence of a large information set where fully parametric models often produce more accurate forecasts of downside risks \citep{CCM2020BVAR}. As a matter of fact, when using quantile regression, including multiple lags of the dependent and independent variables so as to capture the rich autocorrelation structure of macroeconomic and financial time series becomes very impractical and often leads to imprecise estimates of the coefficients and problems such as quantile crossing. Moreover, the entire predictive distribution of the target variables can only be obtained in two steps by interpolating the estimated quantiles with a flexible distribution.  In the light of these limitations a new wave of studies have recently brought some evidences in favour of the use of fully parametric models to assess and predict tail risk to macroeconomic outcomes. \citet{BROWNLEES2021312} for example, show that standard GARCH models have superior forecasting performance with respect to quantile regression methods for forecasting downside risk to GDP growth. As well, \citet{CCM2020BVAR} show that a Bayesian VAR with stochastic volatility performs comparably to quantile regression for estimating and forecasting tail risks. Here I follow and extend this line of research by considering fully parametric models featuring both time varying volatility and time varying skewness, as recently done by \citet{petrelladellemonache2021}, \citet{iseringhausen2021time}, \citet{wolf2021estimating} and \citet{montes2022skewed}. While the first three contributions are all univariate \footnote{\citet{petrelladellemonache2021}  propose a score driven model with \textit{Skew-t} innovations. \citet{ISERINGHAUSEN2020} is the first paper to introduce time varying conditional skewness in a univariate stochastic volatility model by exploiting a \textit{Noncentral-t} distribution for the innovations. \citet{wolf2021estimating} exploits the \textit{Skew-Normal} distributions  but considers a different parametrization for the shocks with respect to the univariate model that I consider in Section \ref{sec:univar} relying as well on a different estimation strategy.} in this paper I model time varying volatility together with time varying skewness both in a univariate and in a multivariate framework. The main advantages of the multivariate framework is that it allows to jointly model the dynamic relationship between the target variables and the risk factors and to explicitly model tail risk to multiple macroeconomic outcomes of interest. The multivariate model that I propose in this paper is a  VAR  model in which Bayesian shrinkage can be conveniently used to avoid over-fitting when exploiting a potential large information set due both to the inclusion of larger number of macroeconomic variables and of a meaningful number of lags needed to properly account for the rich autocorrelation structure of the macroeconomic and financial time series. The model features two distinct stochastic processes respectively governing the time varying volatility and the time varying skewness of the shocks. By considering distinct stochastic processes for the skewness and the volatility of the shocks, this model is different from the Bayesian VAR with \textit{Skew-Normal} shocks introduced by \citet{montes2022skewed} where the latent stochastic process governing the shape of the shocks influences not only the conditional skewness, but also the conditional mean and the conditional variance of the variables in the system. As well, the model differs from \citet{KARLSSON} who recently proposed  a general class of generalized hyperbolic skew Student’s distribution with stochastic volatility for the shocks of the VAR in which the time variation in the volatility of the shocks drives also time variation in their skewness. To my knowledge, this is the first paper that estimates a VAR with two distinct stochastic processes for the volatility and the skewness of the shocks. \\

\textbf{Outline} \hspace{0.2cm} The rest of the article is organized as follows. In Section  \ref{sec:univar} I present the univariate stochastic volatility models with  \textit{Skew-Normal} and \textit{Skew-t} shocks featuring time varying skewness. Then in Section \ref{sec:VAR} I exploit the same conceptual framework to model time varying skewness together with time varying volatility in the shocks of a VAR model.  In both sections I present posterior simulation samplers used for Bayesian estimation of these models. In Section \ref{sec:gar} I and use the models to predict downside risk to GDP growth and compare the forecasting performances to the popular two step approach based on quantile regression by \citet{ABG}. Finally, in Section \ref{sec: medscale} I estimate a medium scale VAR model and show that many macroeconomic and financial variables exhibit time varying conditional skewness. \\

\section{Models}\label{sec:models}
%On the top of that, in order to model time variation, I specify  In particular, to capture asymmetries and heavy tails, I consider both the 
%
\subsection{Univariate time varying skewness stochastic volatility model}\label{sec:univar}
Stochastic volatility models currently represent the state of the art for modelling and forecasting macroeconomic and financial time series.  The basic stochastic volatility model of \citet{jpr1993} specifies a \textit{log-normal} auto-regressive process for the conditional variance with independent innovations in the conditional mean and conditional variance equation. In a second contribution, \citet{JPR2004} introduce a stochastic volatility model that features correlation between the volatility and mean innovations (\textit{leverage effects}) allowing for conditional skewness, but without modelling it explicitly.  \citet{CappuccioLubianRaggi2004} present a stochastic volatility model where the shocks feature a \textit{Skew-GED} distribution while \citet{A2015} introduce a stochastic volatility with \textit{Skew-t} innovations. Both contributions explicitly model conditional skewness, but do not allow for time varying conditional skewness.
Here I present a direct extension of the univariate stochastic volatility model of \citet{jpr1993} that instead explicitly allows for time varying conditional skewness. 

In order to model asymmetries in the conditional distribution of the dependent variable, I assume that the innovations in an otherwise standard stochastic volatility model follow a potentially asymmetric distribution, being the \textit{Skew-Normal} \citep{azzalini1986} and the \textit{Skew-t} \citep{AC2003} distribution. The $Skew-Normal(\zeta,\omega^2,\lambda)$ is an asymmetric distribution fully characterized by three parameters: the location parameter $\zeta$, the scale parameter $\omega^2$ and the shape parameter $\lambda$. The shape parameter $\lambda$ governs the skewness of this distribution. As $\lambda = 0$ the \textit{Skew-Normal} becomes symmetric and collapses to the \textit{Normal}. Positive values of $\lambda$ are associated with a right skewed distribution while negative values of $\lambda$ are associated with a left skewed distribution. \footnote{See Appendix \ref{sec:appendixA} for details on  the \textit{Skew-Normal and \textit{Skew-t}}.}  
To model time variation in the shape of the shocks, I treat the shape parameter $\lambda$ as an additional stochastic process in the model: %    
\begin{equation}\label{unimdl}
    y_t = \boldsymbol{x_{t}\pi} + \sqrt{h_t}\varepsilon_t \hspace{2cm}  \varepsilon_t \sim \text{\textit{Skew-Normal}}(\zeta_{t}, \omega_{t}^2,\lambda_{t})
\end{equation}
\begin{equation}\label{logv}
    log(h_t) = \phi_hlog(h_{t-1}) + \eta_t \hspace{2.5cm} \eta_{t} \sim \mathcal{N}(0, \sigma^2_{\eta}) 
\end{equation}
\begin{equation}\label{lambdat}
    \lambda_{t} = \phi_{\lambda}\lambda_{t-1} + \xi_{t}   \hspace{4cm} \xi_{t} \sim \mathcal{N}(0, \sigma^2_{\xi}) 
\end{equation}
where $y_t$ is the dependent variable observed over the periods $t=1,\ldots,T$, while $\boldsymbol{x_{t}}$ is a row vector of that might contain lags of the dependent variable and other exogenous regressors and $\boldsymbol{\pi}$ is the column vector of coefficients.
%
%This is an otherwise standard stochastic volatility model where the innovations follow the \textit{Skew-Normal} distribution \citep{azzalini1986}.
I assume that the \textit{Skew-Normal} shocks have zero mean and unit variance, that is $\E[\varepsilon_t]= 0$ and $var(\varepsilon_t) = 1 $, which implies the following constraints on the location and scale parameters: 
\begin{equation}\label{zeta}
    \zeta_{t}  =   -\omega_{t}\delta_{t}\sqrt{\frac{2}{\pi}}  \quad \quad \quad   \lor t    
\end{equation} 
\begin{equation}\label{omega}
    \omega^2_{t}=  \left[  1 - \frac{2}{\pi}\delta_{t}^2 \right]^{-1} \quad \quad  \lor t 
\end{equation}
where $\delta_t = \frac{\lambda_t}{\sqrt{1 + \lambda_t^2}}$, with $-1<\delta_t<1$. This parametrization ensures that $\E[y_t |\mathcal{I}_{t-1}] = \boldsymbol{x_{t}\pi} $. In this regard, it is important to remark that imposing $\zeta_t = 0$ instead of (\ref{zeta}) would imply $\E[\varepsilon_t ] \neq 0 $, and in general $\E[\varepsilon_t|\mathcal{I}_{t-1}] \neq 0$.\footnote{ Imposing $\zeta_t = 0$ instead of (\ref{zeta}) leads to a model with a time varying intercept, shifting the conditional mean of $y_t$ proportionally to $\lambda_{t-1}$.} As well, this parametrization ensures that $y_t$ features both time varying conditional volatility and time varying conditional skewness with the former exclusively governed by the stochastic process in equation (\ref{logv}) while the latter by the stochastic process in (\ref{lambdat}).\footnote{It is possible to have a model that features both time varying volatility and time varying skewness by assuming: 
\begin{equation*}
 \begin{array}{l}
 y_t = \boldsymbol{x_{t}\pi} + \varepsilon_t \hspace{2cm}  \varepsilon_t \sim \text{\textit{Skew-Normal}}(\zeta_{t}, \omega_{t}^2,\lambda_{t}) \\
 \lambda_{t} = \phi_{\lambda}\lambda_{t-1} + \xi_{t}   \hspace{4cm} \xi_{t} \sim \mathcal{N}(0, \sigma^2_{\xi}) 
\end{array}
\end{equation*}
assuming $\mathbb{E}(\varepsilon_t) = 0$ (hence (\ref{zeta}) still holds) and imposing $\omega^2 = 1$ which implies  $var(\varepsilon_t) \neq 1 = \left( 1 - \frac{2\delta_t^2}{\pi} \right)$. However, in this case the parameter $\lambda_t$ would drive both conditional skewness and conditional volatility. This is not desirable in general, since we might want to model these two distinct features using different dynamics. } 

In order to explicitly model heavy-tails, together with time-varying skewness, I also consider an alternative specification where the innovations are distributed as a ${Skew-t(\zeta_{t}, \omega_{t}^2,\lambda_{t},\nu)}$ \citep{AC2003}. The parameter of the degrees of freedom $\nu$ determines the tail thickness of the $\textit{Skew-t}$ distribution: as $\nu \rightarrow \infty $ the  $\textit{Skew-t}$  converges to the \textit{Skew-Normal} while when $\lambda = 0$  the \textit{Skew-t}  collapses to a \textit{Student-t}  with $\nu$ degrees of freedom. In this case the constraints on the location and scale parameters that ensure $\E[\varepsilon_t]= 0$ and $var(\varepsilon_t) = 1 $ become:
\begin{equation}
\zeta_{t} = - \omega_{t} \delta_{t} k_1\sqrt{\frac{2}{\pi}} \quad \quad \quad   \lor t
\end{equation}
\begin{equation}
 \omega^2_{t} = \left(  k_2 - \frac{2}{\pi}k_1^2\delta_{t}^2 \right)^{-1}  \quad \quad \quad   \lor t 
\end{equation}
where $k_1 = \sqrt{\frac{\nu}{2}}\frac{\Gamma(\frac{\nu-1}{2})}{\Gamma(\frac{\nu}{2})}$,  $k_2 = \frac{\nu}{\nu - 2}$ and $\Gamma(.)$ is the Gamma function.  This stochastic volatility model with \textit{Skew-t} shocks includes as special cases both the stochastic volatility model with heavy tails without conditional skewness of \citet{JPR2004} and the model with heavy tails and constant conditional skewness of \citet{A2015}.\footnote{The stochastic volatility model with heavy tails of \citet{JPR2004} is a particular version of this model where the shape parameter is constant and equal to 0, that is $\lambda_t = 0 \hspace{0.2cm} \lor t $.  As well, the stochastic volatility model with skewness and heavy tails of is a particular version of this model where $\sigma^2_{\xi} \rightarrow 0 $ and $\phi_{\lambda} = 1$, namely  the shape parameter $\lambda_t$ is constant.} It is straightforward to modify this specification by assuming a different dynamics for the log-volatility and the shape parameter in the state equations (\ref{logv}) and (\ref{lambdat}). For example if we suspect that some of the variables in $\boldsymbol{x_{t}}$ affect not only the conditional mean, but also the conditional variance and the conditional skewness of $y_t$, we can include them in the state equations of these two distinct stochastic processes. For example, as it will be shown in the application to the Growth at Risk framework in Section \ref{sec:gar}, motivated by the findings of \citet{ABG} and subsequent work by \citet{petrelladellemonache2021}, \citet{montes2022skewed} and \citet{wolf2021estimating} I consider a specification in which financial condition affect not only the conditional mean but also the conditional skewness of the future GDP growth distribution.

\subsubsection{Priors and estimation of the univariate TVSSV model} 
This section develops a posterior simulation sampler which allows for Bayesian estimation of the univariate models presented above. For what concerns the specification of the prior distribution for the parameters of the model, I assume a \textit{Normal} prior for the regression coefficients ($\boldsymbol{\pi}$) and for the coefficients in the state equations ($\phi_{\lambda}$ and $\phi_{h}$) while  I specify an Inverse Gamma Prior for the variances of the innovations to the log-volatility and to the shape parameter ($\sigma^2_{\eta}$ and $\sigma^2_{\xi})$. The estimation strategy leverages on the fact that  $\varepsilon_{t} \sim \text{\textit{Skew Normal}}(\zeta_{t},\omega^2_{t},\lambda_{t})$ has the following stochastic representation :  
\begin{equation}\label{repsn}
    \varepsilon_{t} =  \zeta_{t} + \delta_{t}\omega_{t}v_{t} + \sqrt{(1-\delta_{t}^2)}\omega_{t}z_{t}
\end{equation}
%
%that conditionally on the mixing variable $ v_{t}$, the random variable $ \varepsilon_{t}$ is distributed as a Normal.
where $v_{t} \stackrel{i.i.d}{\sim} \text{Truncated Normal}_{[0,\infty)}(0,1)$ and $z_{t} \stackrel{i.i.d}{\sim} \mathcal{N}(0,1)$. Equation (\ref{repsn}) implies that conditioning on the mixing variable $v_{t}$ and on $\delta_t$, which is one to one map to $\lambda_t$, the random variable $\varepsilon_{t}$ is distributed as a \textit{Normal} . This result greatly simplifies the derivation of the full conditional  distributions in the \textit{Gibbs Sampler} and allows to exploit and adapt many of the results used for the estimation of the standard stochastic volatility model with Gaussian innovations \citep{jpr1993}. In particular, in the model with \textit{Skew-Normal} shocks, once I have obtained a draw from the full conditional posterior distribution of the mixing variable $v_t$ and from the full conditional distribution of the shape parameter $\lambda_t$, I can exploit the conditionally \textit{Normal} distribution of $\varepsilon_t$ in the derivation of formulas of the conditional distributions of the other parameters and the latent states of the model. 
%defining  $u_t \equiv \sqrt{h_t}\varepsilon_t$
%\begin{equation}\label{gausskernel}
% y_t|. \sim \mathcal{N}\left(  \boldsymbol{x_{t}\beta}   + \sqrt{h_t}(\zeta_t + \omega_t\delta_t v_t), h_t\omega^{2}_t(1 - \delta_t^2) \right)   
%\end{equation}
%
%This equation implies that  \\
%
%\begin{equation}\label{media_u}
% \mu_{u} =  \sqrt{h_t}(\zeta_t + \omega_t\delta_t v_t)    
%\end{equation}
%\begin{equation}\label{var_u}
%  \sigma_u^2 =  h_t\omega^{2}_t(1 - \delta_t^2)   
%\end{equation}
%As a matter of fact, once I draw from the full conditional posterior $p(v_{1} \ldots, {v_{T}}|\boldsymbol{\Theta}, \boldsymbol{\lambda}), \boldsymbol{h})$ of the parameters and of the states $p(h_{1} \ldots, {h_{T}}|\boldsymbol{\Theta}, \boldsymbol{\lambda}), \boldsymbol{v})$ $p(\lambda_{1} \ldots, {\lambda_{T}}|\boldsymbol{\Theta}, \boldsymbol{h}), \boldsymbol{v})$ \\
 % $\sigma_{\eta}^2 \sim IG \left( \nu_{\sigma_{\xi}^2}, s_{\sigma_{\eta}^2} \right)$ and $\sigma_{\xi,i}^2 \sim IG \left( \nu_{\sigma_{\xi}^2}, s_{\sigma_{\xi}^2} \right)$. \\
%where $\Sigma_{\beta}$ has a Minnesota type structure
% 
Moreover $\zeta_t$, $\omega_t$ and $\delta_t$
are neither parameters nor latent states to be estimated. $\zeta_t$ and $\omega_t$  satisfy the constraints  (\ref{zeta}) and (\ref{omega}) and ensure the correct parameterization of the shocks at each time period $t = 1, \ldots, T$, while $\delta_t$ is a one to one map to $\lambda_t$, namely $\delta_t = \frac{\lambda_t}{\sqrt{1 + \lambda_t^2}}$.  

Table \ref{tab:Gibbs sampler} presents the details on the \textit{Gibbs Sampler} while Appendix \ref{sec:appendixB} reports the derivations of the full conditional posterior distributions. In  Step 1) I sample the mixing variables $\{v_{t}\}_{t=1}^T$ from the full conditional posterior  distribution $p(v_{t}|\boldsymbol{\Theta}, \boldsymbol{\lambda}, \boldsymbol{h}, \boldsymbol{y})$ which is a \ \textit{Truncated Normal} distribution.   Steps 2) 3) 4) 5) 6) are pretty standard: I draw the regression coefficients $\boldsymbol{\pi}$ in the observation equation (\ref{unimdl}) and the autoregressive coefficients and the variances in the two state equations (\ref{logv}) (\ref{lambdat}) from their respective full conditional posterior distributions. In Step 7) and Step 9) I draw the initial states for the volatility $h_0$ and the shape parameter $\lambda_0$, while in Steps 8) and 10) I draw the entire history for the volatilities and the shape parameters. Since it is not feasible to directly sample from  the full conditional  distributions of the volatilities $p(h_{1}, \ldots, h_{T}| \boldsymbol{\Theta},\boldsymbol{v}, \boldsymbol{\lambda}, \boldsymbol{y})$ and the shape parameters $p(\lambda_{i1}, \ldots, \lambda_{iT}| \boldsymbol{\Theta}, \boldsymbol{v}, \boldsymbol{h}, \boldsymbol{y}) $ I rely on the particle filter to approximate these distributions. In alternative to the particle step, to draw both the log-volatilities and the shape parameters it is possible to consider an independence Metropolis Hastings step but I experienced that the algorithm based on the particle filter has smaller mixing times.\footnote{In the particle steps, in order to alleviate path degeneracy, I exploit the \textit{Ancestor Sampling} procedure developed in \citet{lindsten2014particle} which enables fast mixing even when using seemingly few particles. \citet{lindsten2014particle} study the properties of the sampler and provide the formal proof for the convergence of the algorithm.}  In the particle approximation, I use the transition equations (\ref{logv}) and (\ref{lambdat}) as importance densities and compute the weights accordingly.  The details on the particle steps used to approximate the full conditional posterior distribution of the volatilities and the shape parameters can be find in Table \ref{tab:part} in the Appendix \ref{APSTEP}. As well, in the Appendix \ref{APSTEP}, I report the details on the steps of the alternative algorithm which relies on the independence Metropolis Hastings steps to draw the volatilities and the shape parameters. 
\begin{table}[H]
\begin{center}
\caption{MCMC algorithm for the univariate TVSSV model}
\label{tab:Gibbs sampler}
\begin{tabular}{l l}
\hline 
\small \textit{MCMC for the univariate TVSSV model}  &  \\  
\hline\hline
\scriptsize Initialize $\boldsymbol{\Theta}^{(0)}, \boldsymbol{s}^{(0)}$\\
\scriptsize For $m = 0 : \text{Total MCMC draws}$                       \\
\scriptsize \hspace{1cm} 1) Draw $\{v_{t}\}_{t=1}^{T^{(m+1)}} $ from $p(v_{1} \ldots, {v_{T}}|\boldsymbol{\Theta}^{(m)}, \boldsymbol{\lambda}^{(m)}), \boldsymbol{h}^{(m)}, \boldsymbol{y})$ \\
\scriptsize \hspace{1cm} 2) Draw $\boldsymbol{\pi}^{(m+1)}$ from $p(\boldsymbol{\pi}| \boldsymbol{\Theta}^{(m)}, \boldsymbol{v}^{(m)}, \boldsymbol{\lambda}^{(m)}), \boldsymbol{h}^{(m)}, \boldsymbol{y}) $ \\
\scriptsize \hspace{1cm} 3) Draw $\sigma_{\eta}^{2^{(m+1)}}$ from $p (\sigma^2_{\eta}|\boldsymbol{\Theta^{(m)},v^{(m)}}, \boldsymbol{\lambda}^{(m)}), \boldsymbol{h}^{(m)}, \boldsymbol{y}) $   \\
\scriptsize \hspace{1cm} 4) Draw $\sigma_{\xi}^{2^{(m+1)}}$ from $p (\sigma^2_{\xi}|\boldsymbol{\Theta^{(m)},v^{(m)}}, \boldsymbol{\lambda}^{(m)}), \boldsymbol{h}^{(m)}, \boldsymbol{y}) $   \\
\scriptsize \hspace{1cm} 5) Draw $\phi_h^{(m+1)}$ from $p (\phi_{h}|\boldsymbol{\Theta^{(m)},v^{(m)}}, \boldsymbol{\lambda}^{(m)}), \boldsymbol{h}^{(m)}, \boldsymbol{y}) $  \\
\scriptsize \hspace{1cm} 6) Draw $\phi_{\lambda}^{(m+1)}$ from $p (\phi_{\lambda}|\boldsymbol{\Theta^{(m)},v^{(m)}}, \boldsymbol{\lambda}^{(m)}), \boldsymbol{h}^{(m)}, \boldsymbol{y}) $            \\
\scriptsize \hspace{1cm} 7) Draw $h_{0}^{(m+1)}$ from $p(h_{0}| \boldsymbol{\Theta}^{(m)}, \boldsymbol{v}^{(m)},  \boldsymbol{\lambda}^{(m)}), \boldsymbol{h}^{(m)}, \boldsymbol{y})$ \\
\scriptsize \hspace{1cm} 8) Draw $\{h_{t}\}_{t=1}^{T^{(m+1)}}$ from $p(h_{1}, \ldots, h_{T}| \boldsymbol{\Theta^{(m)}}, \boldsymbol{v^{(m)}}, \boldsymbol{\lambda}^{(m)}, \boldsymbol{y}) $  \\
\scriptsize \hspace{1.2cm} \uline{Particle Step} \\
\scriptsize \hspace{1cm} 9) Draw $\lambda_{0}^{(m+1)}$ from $\lambda_{0}^{(m+1)}$ from $p(\lambda_{0}| \boldsymbol{\Theta}^{(m)}, \boldsymbol{v}^{(m)},\boldsymbol{\lambda}^{(m)}, \boldsymbol{h}^{(m)}, \boldsymbol{y})$    \\
\scriptsize \hspace{1cm} 10) Draw $\{\lambda_{t}\}_{t=1}^{T^{(m+1)}}$ from $p(\lambda_{i1}, \ldots, \lambda_{iT}| \boldsymbol{\Theta^{(m)}}, \boldsymbol{v^{(m)}}), \boldsymbol{h}^{(m)}, \boldsymbol{y}) $                      \\
\scriptsize \hspace{1.2cm} \uline{Particle Step} \\
\scriptsize end \\ 
\hline\hline
\end{tabular}
\end{center}
\end{table}
%
%where $s_t$ stands for the generic unobserved latent state being $log(h_t)$ in Step 5) and $\lambda_t$ in Step 7) of the Gibbs Sampler in Table \ref{tab:Gibbs sampler}. As anticipated above a valid particle approximation to the Gibbs Sampler requires a Conditional Sequential Monte Carlo update which guarantees that a pre-specified path of the state variables is ensured to survive all the resampling steps \citep{andrieu2010particle}. Hence, if I consider a generic iteration $ m + 1 $ of the Gibbs Sampler, when using $K$ particles to approximate $p(h_{1}, \ldots, h_{T}| \boldsymbol{\Theta},\boldsymbol{v}, \boldsymbol{\lambda})$ and $p(\lambda_{i1}, \ldots, \lambda_{iT}| \boldsymbol{\Theta}, \boldsymbol{v}, \boldsymbol{h}) $ ,  only $K-1$ particles are generated while the $K^{th}$ particle is set to the pre-specified path $ h_{1:T}^{(m)}$ and $ \lambda_{1:T}^{(m)}$. %namely: 
%\begin{equation}
%    w_{t}^k = f(y_t|s_t, .)\hspace{1cm} \text{for} \hspace{0.2cm} k = 1,\ldots, K
%\end{equation}
%
%where $ f(y_t|s_t, .)$ is given the Gaussian distribution, obtained by conditioning on the mixing variables, the parameters and the other states in the model. 
%by equation (\ref{gausskernel}). 
%We refer to the original paper, \citet{lindsten2014particle} for the details on the \textit{Ancestral Sampling} step, that for $t>2$ artificially assign a history to the partial pre-specified path  $s_{t:T}^{(m)}$.   
To estimate the version of the model with \textit{Skew-t} innovations, I just exploit the fact that $\varepsilon_{t} \sim \text{\textit{Skew-t}}(\zeta_{t},\omega^2_{t},\lambda_{t}, \nu)$ has in turn a convenient stochastic representation, namely:  
\begin{equation}\label{rapst}
    \varepsilon_{t} =  \zeta_{t} + \delta_{t}\omega_{t}o_{t}^{-0.5}v_{t} + \sqrt{(1-\delta_{t}^2)}\omega_{t}o_{t}^{-0.5}z_{t}
\end{equation}
$v_{t} \stackrel{i.i.d}{\sim} \text{Truncated Normal}_{[0,\infty)}(0,1)$, $z_{t} \stackrel{i.i.d}{\sim} \mathcal{N}(0,1)$ and $o_{t} \stackrel{i.i.d}{\sim} \mathcal{G}(\frac{\nu}{2},\frac{\nu}{2})  $ \\
%
%In this case, conditioning on the two mixing variables $v_t$ and $o_t$, the random variable $\varepsilon_t$ is distributed as a Normal. 
%
This is the same same representation of the \textit{Skew-Normal} except for the additional mixing variable $o_t$. Therefore, conditioning on both the two mixing variables $m_t = \{v_{t}, o_{t} \}$ and on $\delta_{t}$, which is a one to one map with $\lambda_{t}$, the shock $\varepsilon_t$ is distributed as a \textit{ Normal}. Therefore, also in this case, I can exploit and adapt the derivations of the standard model with Gaussian shocks when deriving the full conditional posterior distribution in the \textit{Gibbs Sampler}.  
%
%\begin{equation}
% y_t|. \sim \mathcal{N} \left(  \boldsymbol{x_{t}\beta} + \sqrt{h_t}(\zeta_t + \omega_t\delta_t o_t^{-0.5}, h_t\omega^{2}_t(1 - \delta_t^2) o_t^{-1} \right)   
%\end{equation}
%
In order to estimate the model it is just needed to consider a further initial step to draw from $p(o_{1} \ldots, {o_{T}}|\boldsymbol{\Theta}, \boldsymbol{v},\boldsymbol{\lambda}, \boldsymbol{h}, \boldsymbol{y})$, namely:
\footnote{ In the estimation of the model with heavy tails (\textit{Skew-t} shocks), I fix the tail thickness parameters $\nu$ to 5. Given the relative short time series length of macroeconomic data, it is particularly difficult to make inference on this parameter. In general, you can draw this parameter adding another Metropolis Hastings step to draw from  $p(\nu|\boldsymbol{\Theta}, \boldsymbol{v}, \boldsymbol{o},\boldsymbol{\lambda}, \boldsymbol{y})$} 
\begin{table}[H]
\begin{tabular}{l l}
\scriptsize \hspace{4cm}  Draw $\{o_{t}\}_{t=1}^{T^(m+1)} $ from $p(o_{1} \ldots, {o_{T}}|\boldsymbol{\Theta}^{(m)}, \boldsymbol{v}^{(m)}, \boldsymbol{\lambda}^{(m)}), \boldsymbol{h}^{(m)}, \boldsymbol{y})$ 
\end{tabular}
\end{table}
and then adapt Steps 2) to 10) in Table (\ref{tab:Gibbs sampler}) with the new formulas of the full conditional distributions derived by conditioning on the further mixing variables $\{o_{t}\}_{t=1}^T$.  In this case, since it is not possible to directly sample from the full conditional distribution of the mixing variable $o_t$, I use Metropolis Hastings to simulate draws from this distribution. Appendix \ref{draw} reports the details of this step.

\subsection{Time varying skewness stochastic volatility VAR model}\label{sec:VAR}

Given the risk management nature of the problem of policymaking, it is often the case that the objective of interest is to quantify and predict tail risk to multiple macroeconomic outcomes \citep{km}. In particular, from a modelling perspective, we might be interested in a multivariate model that can characterize asymmetries in the future distribution of multiple macroeconomic time-series. VAR models \citep{sims1980macroeconomics} emerged as the natural tool to capture the rich dynamic interrelationship between multiple macroeconomic time series. They currently represent the workhouse in empirical macroeconomics and are routinely used for forecasting and policy analyses \citep{SW2012}. In this section I exploit the conceptual framework presented in the previous section to jointly model the dynamic  behaviour of multiple time series in a Bayesian VAR model and capture time varying skewness in the conditional distribution of the variables in the system. The model is given by:
\begin{equation}\label{model}
   \boldsymbol{y_t} = \boldsymbol{\Pi_0}  + \boldsymbol{\Pi_1y_{t-1}} + \ldots + \boldsymbol{\Pi_py_{t-p}} +  \boldsymbol{A}^{-1}\boldsymbol{H_t}^{0.5}\boldsymbol{\varepsilon_t}   
\end{equation}

where $\boldsymbol{y}_t$ is an  $N \times 1$ vector of variables observed over the periods $t=1,\ldots,T$.  $\boldsymbol{H_t}$ is a diagonal matrix that contains the volatilities on its main diagonal, namely ${\boldsymbol{H_t} = diag(h_{1,t} \ldots, h_{N,t})}$ and $\boldsymbol{A}^{-1}$ is a lower triangular matrix with ones on its main diagonal. The log-volatilities evolve over time according to: 
\begin{equation}\label{loghti}
 log(h_{i,t}) = \phi_{h,i}log(h_{i,t-1}) + \eta_{i,t}     \hspace{3cm} \eta_{i,t} \sim N(0, \sigma^2_{\eta,i})
\end{equation}
for  $i=1,\ldots,N$. In the Gaussian stochastic  volatility model of \citet{COGLEY2005262} and \citet{primiceri2005} it is assumed $\boldsymbol{\varepsilon_t} \sim N(0,\boldsymbol{I})$. In our specification, $\boldsymbol{\varepsilon_t}$ is a vector of \textit{Skew-Normal} shocks, namely:
\begin{equation}\label{sn}
    \boldsymbol{\varepsilon_t} = [\varepsilon_{1t}, \ldots, \varepsilon_{Nt}]' \hspace{2cm} \varepsilon_{it} \sim \text{\textit{Skew-Normal}}(\zeta_{it}, \omega_{it}^2,\lambda_{it})
\end{equation} 
where the shape parameters $\lambda_{it}$ evolve according to: 
\begin{equation}\label{lambasi}
 \lambda_{i,t} = \phi_{\lambda,i}\lambda_{i,t-1} + \xi_{i,t}     \hspace{4cm}  \xi_{i,t} \sim N(0, \sigma^2_{\xi,i})
 \end{equation}
%
%Defining $\boldsymbol{u}_t = \boldsymbol{A^{-1}H_t^{0.5}\varepsilon}_{t}$, equations (\ref{zeta}) and (\ref{omega})  together with the assumption of independence of the \textit{Skew-Normal} shocks $\varepsilon_{it}$ imply the usual factorization of the variance covariance matrix of the innovations: 
%
%\begin{equation}
%var(\boldsymbol{u_t}) \equiv \boldsymbol{\Sigma_t} =  \boldsymbol{A}^{-1}\boldsymbol{H_tA^{-1}'}   
%\end{equation}
%
In order to have $\E[\boldsymbol{\varepsilon_t}]=  \boldsymbol{0} $ and $var(\boldsymbol{\varepsilon_t})= \boldsymbol{I} $ the shocks are parameterized imposing the constraints on the location parameters $\zeta_{it}$ and on the scale parameters $\omega_{it}$ discussed in the previous section. As in the univariate framework, I can explicitly model heavy-tails, together with time-varying skewness, by considering an alternative specification where: 
\begin{equation}\label{st}
    \boldsymbol{\varepsilon_t} = [\varepsilon_{1t}, \ldots, \varepsilon_{Nt}]' \hspace{2cm} \varepsilon_{it} \sim \text{\textit{Skew-t}}(\zeta_{it}, \omega_{it}^2,\lambda_{it},\nu)
\end{equation} 
The model nests the constant coefficients version of the popular VAR model with stochastic volatility introduced by \citet{COGLEY2005262} and \citet{primiceri2005} and considered in \citet{carriero2019large}.\footnote{As well, the stochastic volatility VAR with fat tails in \citet{CLARKRAV2015} is also a special case of this model with $\lambda_{i,t} \hspace{0.2cm} = 0 \lor i,t$. \citet{KARLSSON}  stochastic volatility model VAR with \textit{Skew-t} orthogonal residual is as well a particular version of this model with $\phi_{\lambda_i} = 1$ and $\sigma^2_{\xi,i} \rightarrow 0 \hspace{0.2cm} \lor i$.} In these models, as long as the short run restrictions implied by the Cholesky ordering are satisfied, the shocks can be interpreted as structural.\footnote{It is worth to mention that due to the ``Cholesky type" specification of the stochastic volatility VAR model considered here, the order in which the variables enter in the VAR matters not only for the identification of the shocks but also for the estimation of the model. This is fact was stressed first by \citet{primiceri2005} and more recently by \citet{shin2021} and \citet{chan2021large}. On the lines of the work of \citet{chan2021large} I am currently working on a order invariant version of the model considered in this paper.} This means that, other than for forecasting purposes, the model can be practically used for policy analysis and structural scenario analyses. %This framework extends the familiar VAR framework to account for time-varying asymmetries in the predictive distribution of multiple time series. 
Also in this multivariate framework it is straightforward to modify the specification of the state equations of the log-volatilities  and the shape parameters by assuming a different dynamics in (\ref{loghti}) and (\ref{lambasi}). For example, as it will be shown in the empirical application in Section \ref{sec:gar}, I can capture the nonlinear relationship between two variables in the VAR  by including the lags of one variable in the state equations of the log-volatility and/or the shape parameter of the shocks to the other variable. 

%Gaussian stochastic volatility VAR models are typically estimated through Monte Carlo Markov Chain algorithm  that has the desired distribution   in a Bayesian setting which combines the Gaussian distribution coming from the likelihood with the priors of the parameters and the states \citep{COGLEY2005262,primiceri2005}. \\ 

\subsubsection{Priors and estimation of the TVS-SV VAR}
For what concerns the choice of the prior distributions for the parameters of the model, I assume a Normal prior for the autoregressive coefficients $vec(\boldsymbol{\Pi})$. As well, following \citet{COGLEY2005262}, I specify a Normal prior for the free
elements in the matrix $\boldsymbol{A}$. Finally, as in the univariate framework, I specify independent Inverse Gamma priors for the variance of the innovations to the log-volatilities and to the shape parameters ($\sigma^2_{\eta,i}$ and $\sigma^2_{\xi,i})$ and Normal priors for the coefficients in the state equations ($\phi_{h,i}$ and $\phi_{\lambda,i})$.  
The estimation strategy for the VAR model is just a generalization of the one for the univariate model that again leverages on the stochastic representation of the \textit{Skew Normal}  (\ref{repsn})  and \textit{Skew-t} (\ref{rapst}) shocks. Exploiting this representation, I can write the vector of \textit{Skew-Normal} shocks $\boldsymbol{\varepsilon_t}$ as follows: \footnote{Note that the powers on the matrices refer all to diagonal matrices. For example $(\boldsymbol{I}_N - \boldsymbol{\Delta_t}^2) = diag( \sqrt{1 - \delta_{1,t}^2}, \ldots, \sqrt{1 - \delta_{N,t}^2}  )$ or afterwords
$\boldsymbol{O_t}^{-0.5} = diag\left(\frac{1}{\sqrt{o_{1,t}}}, \ldots, \frac{1}{\sqrt{o_{N,t}}}\right)$ }
\begin{equation}
\boldsymbol{\varepsilon_t} =  \boldsymbol{\zeta_t} +  \boldsymbol{\Omega_t} \boldsymbol{\Delta_t} \boldsymbol{v_t} + {\boldsymbol{\Omega_t}}(\boldsymbol{I}_N - \boldsymbol{\Delta_t}^2)^{0.5}\boldsymbol{z_t} 
\end{equation}
where: 

$\boldsymbol{\zeta_t} =  [\zeta_{1,t}, \ldots, \zeta_{N,t}]'$

$\boldsymbol{\Omega_t} =  diag( \omega_{1t} \ldots \omega_{Nt})$ 

$\boldsymbol{\Delta_t} =  diag( \delta_{1t} \ldots \delta_{Nt})$ 

$\boldsymbol{v_t} =  [v_{1,t}, \ldots, v_{N,t}]' \hspace{2cm}  v_{i,t} \sim TruncatedNormal_{(0,\infty)}(0,1)$

$\boldsymbol{z_t} = [z_{1,t}, \ldots, z_{N,t}]' \hspace{2cm}  z_{it} \sim  N(0,1)$. \\

%\begin{equation}\label{repvar}
% \boldsymbol{A}\boldsymbol{u}_t = \boldsymbol{H}_t^{0.5} [\boldsymbol{\zeta_t} +  \boldsymbol{\omega_t} \boldsymbol{\Delta_t} \boldsymbol{v_t} + \boldsymbol{\omega_t}(\boldsymbol{I} - \boldsymbol{\Delta_t}^2)^{0.5}\boldsymbol{z_t}]    
%\end{equation}
%
As in the univariate framework, I can exploit this result when deriving the full conditional posterior distributions of the parameters and the unobserved states in the \textit{Gibbs Sampler}. As a matter of fact, also in this case, $\zeta_{it}$ and $\omega_{it}$ respectively stored in the column vector $\boldsymbol{\zeta_t}$ and in the diagonal matrix $ \boldsymbol{\Omega_t}$ are neither parameters nor latent states to be estimated.
$\zeta_{it}$ and $\omega_{it}$ are fixed to
satisfy the constraints  (\ref{zeta}) and (\ref{omega}) and ensure the correct parameterization of the shocks in each equation of the VAR $i = 1, \ldots, N$ and at each time period $t = 1, \ldots, T$. As well, the elements in the diagonal matrix $\boldsymbol{\Delta_t}$ (that is $\delta_{it}$) are one to one map of the latent states $\lambda_{it}$.
%
%
%
%defining $\boldsymbol{u}_t \equiv \boldsymbol{A^{-1}H_t^{0.5}\epsilon_t}$
%\begin{equation}
 %\boldsymbol{y_t|.} \sim \mathcal{N}\left( \sum_{i=1}^p\boldsymbol{\Pi_i}\boldsymbol{y_{t-i}} + %\boldsymbol{A^{-1}H}_t^{0.5} (\boldsymbol{\zeta_t} +  \boldsymbol{\omega_t} \boldsymbol{\Delta_t}  \boldsymbol{v_t}), \boldsymbol{A^{-1}H}_t \boldsymbol{\omega_t^2}(\boldsymbol{I} - \boldsymbol{\delta_t^2)A^{-1}'}\right)    
%\end{equation}
%
% 
%
%
%$\sigma_{\eta,i}^2 \sim IG \left( \nu_{\sigma_{\xi}^2}, s_{\sigma_{\eta}^2} \right)$ and $\sigma_{\xi,i}^2 \sim IG \left( \nu_{\sigma_{\xi}^2}, s_{\sigma_{\xi}^2} \right) \quad i = 1,\ldots, N$.
%

Table (\ref{tab:algorithm}) presents the details of the sampler. In Step 1) I draw the mixing variables $\{v_{it}\}_{t=1}^T$ for $ i=1,\ldots,N$. In Step 2) I draw the coefficients of the VAR coefficients adapting to my framework the correct version of the triangular algorithm developed in \citet{carriero2019large} and corrected in \citet{corrCARRIERO2022}. This approach allows to reduce the computational burden associated to the system-wide estimation of Bayesian VAR  with stochastic volatility and non-conjugate priors by exploiting a triangularization of the system. In Step 3), I adapt the approach of \citet{COGLEY2005262} to draw the free elements in the matrix $\boldsymbol{A}$. In Step 4) 5) and 6) 7) I draw the variances and the autoregressive coefficients of the state equations while in Step 8) and 10) I draw the initial state for the volatilities $h_{i0}$ and the shape parameters $\lambda_{i0}$ . In Step 9) and 11) I draw the entire path for the volatilities and the shape parameters, using the \textit{Particle Step with Ancestor Sampling} described in Table \ref{tab:part} in the Appendix \ref{APSTEP}.

\begin{table}[H]
\begin{center}
\caption{MCMC algorithm for the TVSSV VAR model}
\label{tab:algorithm}
\begin{tabular}{l l}
\hline 
\small \textit{Particle Gibbs Sampler for the TVSSV-VAR model}  &  \\  
\hline\hline
\scriptsize Initialize $\boldsymbol{\Theta}^{(0)}, \boldsymbol{s}^{(0)},\boldsymbol{v}^{(0)}$\\
\scriptsize For $m = 0 : \text{Total MCMC draws}$ \\
\scriptsize \hspace{1cm} 1) Draw $\{ \boldsymbol{v_{it}}\}_{t=1}^{T^{(m+1)}} $ from $p(\boldsymbol{v_{i1}} \ldots, {v_{iT}}|\boldsymbol{\Theta}^{(m)}, \boldsymbol{s}^{(m)}, \boldsymbol{Y})$ & $i = 1,\ldots,N $ \\
\scriptsize \hspace{1cm} 2) Draw $\boldsymbol{\Pi}^{(m+1)}$ from $p(\boldsymbol{\Pi}| \boldsymbol{\Theta}^{(m)}, \boldsymbol{v}^{(m)}, \boldsymbol{s}^{(m)}, \boldsymbol{Y})$ \\
\scriptsize \hspace{1cm} 3) Draw $\boldsymbol{A}^{(m+1)}$ from $p(\boldsymbol{A}| \boldsymbol{\Theta}^{(m)}, \boldsymbol{v}^{(m)}, \boldsymbol{s}^{(m)}, \boldsymbol{Y})$ \\
\scriptsize \hspace{1cm} 4) Draw $\sigma_{\xi,i}^{2^{(m+1)}}$ from $p (\sigma^2_{\xi,i}|\boldsymbol{\Theta^{(m)},s^{(m)},v^{(m)}}, \boldsymbol{Y})$ & $i = 1,\ldots,N $        \\
\scriptsize \hspace{1cm} 5) Draw $\sigma_{\eta,i}^{2^{(m+1)}}$ from $p (\sigma^2_{\eta,i}|\boldsymbol{\Theta^{(m)},s^{(m)},v^{(m)}}, \boldsymbol{Y}) $ & $i = 1,\ldots,N $        \\
\scriptsize \hspace{1cm} 6) Draw $\phi_{h,i}^{(m+1)}$ from $p (\phi_{h,i}|\boldsymbol{\Theta^{(m)},s^{(m)},v^{(m)},Y}) $ & $i = 1,\ldots,N $        \\
\scriptsize \hspace{1cm} 7) Draw $\phi_{\lambda,i}^{(m+1)}$ from $p (\phi_{\lambda,i}|\boldsymbol{\Theta^{(m)},s^{(m)},v^{(m)}}, \boldsymbol{Y}) $ & $i = 1,\ldots,N $        \\       
\scriptsize \hspace{1cm} 8) Draw $h_{i,0}^{(m+1)}$ from $p(h_{i,0}| \boldsymbol{\Theta}^{(m)}, \boldsymbol{v}^{(m)}, \boldsymbol{s}^{(m)}, \boldsymbol{Y}) $ & $i = 1,\ldots,N $\\
\scriptsize \hspace{1cm} 9) Draw $\{h_{it}\}_{t=1}^{T^{(m+1)}}$ from $p(h_{i1}, \ldots, h_{iT}| \boldsymbol{\Theta^{(m)}}, \boldsymbol{v^{(m)}}, \boldsymbol{s}^{(m)}, \boldsymbol{Y}) $   & $i = 1,\ldots,N $ \\
\scriptsize \hspace{1.5cm} \uline{Particle step} \\
\scriptsize \hspace{1cm} 10) Draw $\lambda_{i,0}^{(m+1)}$ from $\lambda_{i,0}^{(m+1)}$ from $p(\lambda_{i,0}| \boldsymbol{\Theta}^{(m)}, \boldsymbol{v}^{(m)}, \boldsymbol{s}^{(m)}, \boldsymbol{Y}) $ & $i = 1,\ldots,N $                    \\
\scriptsize \hspace{1cm} 11) Draw $\{\lambda_{it}\}_{t=1}^{T^{(m+1)}}$ from $p(\lambda_{i1}, \ldots, \lambda_{iT}| \boldsymbol{\Theta^{(m)}}, \boldsymbol{v^{(m)}}, \boldsymbol{s}^{(m)}, \boldsymbol{Y}) $ & $i = 1,\ldots,N $                      \\
\scriptsize \hspace{1.5cm} \uline{Particle step}  \\
\scriptsize end \\ 
\hline \hline
\end{tabular}
\end{center}
\end{table}
\noindent
As in the univariate framework, it is easy to adapt the sampler to a version of the VAR model with \textit{Skew-t} shocks. In this case (\ref{rapst}) becomes: 
\begin{equation}\label{multi_stoc}
\boldsymbol{\varepsilon_t} = \boldsymbol{\zeta_t} +  \boldsymbol{\Omega_t} \boldsymbol{\Delta_t}\boldsymbol{O_t}^{-0.5}  + {\boldsymbol{\Omega_t}}(\boldsymbol{I}_n - \boldsymbol{\Delta_t}^2)^{0.5}\boldsymbol{O_t}^{-0.5}\boldsymbol{z_t} 
\end{equation}
%\begin{equation}
% \boldsymbol{y_t|.} \sim \mathcal{N}\left( \sum_{i=1}^p\boldsymbol{\Pi_i}\boldsymbol{y_{t-i}} + \boldsymbol{A^{-1}H}_t^{0.5} (\boldsymbol{\zeta_t} +  \boldsymbol{\omega_t} \boldsymbol{\Delta_t}\boldsymbol{O_t}^{-0.5}  \boldsymbol{v_t}), \boldsymbol{A^{-1}H}_t \boldsymbol{\omega_t^2}(\boldsymbol{I} - \boldsymbol{\delta_t^2)o_t^{-1}A^{-1}'}\right)    
%\end{equation}
where $\boldsymbol{O_t} =  diag( o_{1t} \ldots o_{Nt}) \hspace{2cm}  o_{it}  \sim  Gamma(\frac{\nu}{2},\frac{\nu}{2})$.\\
It is enough to adapt the Gibbs Sampler  by adding another initial step to draw the mixing variables $\{ o_{it}\}_{t=1}^T$ for $i = 1, \ldots, N$ 

\begin{table}[H]
\begin{tabular}{l l}
\scriptsize \hspace{4cm} Draw $\{o_{it}\}_{t=1}^{T^{(m+1)}} $ from $p(o_{i1} \ldots, {o_{iT}}|\boldsymbol{\Theta}^{(m)}, \boldsymbol{v}^{(m)}, \boldsymbol{s_t}^{(m)}, \boldsymbol{Y})$  & $i = 1,\ldots,N $\\
\end{tabular}
\label{tab:st}
\end{table}
and then to update the formulas of the full conditional posterior distributions in order to account for the extra terms. 
\noindent Again I use Metropolis Hastings to simulate draws from $p(o_{i1} \ldots, {o_{iT}}|\boldsymbol{\Theta}, \boldsymbol{v}, \boldsymbol{s_t})$ for $i = 1, \ldots, N$, since it is not directly possible to sample from these distributions.
\section{Growth at Risk}\label{sec:gar}

The work of \citet{ABG} (henceforth ABG) pioneered a recently growing body of research, which examines the main sources of tail risk to GDP growth in relationship to changes in economic and financial conditions. This section compares the out of sample Growth-at-Risk (GaR) estimates for the U.S from our time varying skewness stochastic volatility models to the two step approach based on quantile regression of ABG .  % The authors find that financial conditions, measured by the \textit{Chicago Fed Index of Financial Conditions} (NFCI) predict changes in the left tail of the future distribution of GDP growth. 
In order to model asymmetric changes in the conditional distribution of GDP growth as a function of changes in financial conditions, I consider the following specification of the univariate TVSSV model: 
\begin{equation}
\begin{array}{l}
gdpgrowth_t = \pi_0 + \pi_1 gdpgrowth_{t-1} + \pi_2 gdpgrowth_{t-2} + \pi_3 NFCI_{t-1} + \sqrt{h}_t\varepsilon_t \\
\hspace{8cm}    \varepsilon_t \sim Skew-Normal( \zeta_t, \omega_t,\lambda_t)
\\ 
\hspace{10cm} \text{or} \\
\hspace{8cm}    \varepsilon_t \sim Skew-t( \zeta_t, \omega_t,\lambda_t,\nu)
\end{array}
\end{equation}
\begin{equation}
log(h_t) = \phi_hlog(h_{t-1}) + \eta_t \hspace{2.5cm} \eta_{t} \sim \mathcal{N}(0, \sigma^2_{\eta}) 
\end{equation}
\begin{equation}\label{skewness_nfci}
    \lambda_t =  \phi_{\lambda}\lambda_{t-1} + \beta_{1} NFCI_{t-1} + \xi_t \hspace{2cm} \xi_t  \sim \mathcal{N}(0,\sigma^2_\xi)
\end{equation}
In this specification the NFCI directly affects the conditional skewness of the future GDP growth distribution. More specifically, the coefficient $\beta_1$ captures changes in the skewness of the conditional distribution of GDP growth as a function of financial conditions. This coefficient is meant to capture the non-linear relationship between deteriorating financial conditions and future GDP growth distribution found in ABG. Since our focus is to model the asymmetric effect of the NFCI on the future GDP growth distribution, I threat the log-volatilities as exogenous autoregressive processes, not affected by the NFCI. As a matter of fact,  augmenting the state equation for the log-volatilities with the NFCI index, as it is done in the state equations of the shape parameters, implies that financial conditions would affect symmetrically both a upper and the lower quantiles of the future GDP growth distribution.  Together with the univariate model, I consider as well a bivariate TVSSV-VAR(2) model where $\boldsymbol{y_t} = [ gdpgrowth, NFCI]'$ and:   
\begin{equation}
\begin{array}{l}
   \boldsymbol{y_t} = \boldsymbol{\Pi_0}  + \boldsymbol{\Pi_1y_{t-1}} +  \boldsymbol{\Pi_2y_{t-2}} +  \boldsymbol{A}^{-1}\boldsymbol{H_t}^{0.5}\boldsymbol{\varepsilon_t}  \\
    \hspace{5 cm} \varepsilon_{it} \sim \text{\textit{Skew-Normal}}(\zeta_{it}, \omega_{it}^2,\lambda_{it}) \\
    \hspace{7cm} \text{or} \\
    \hspace{5cm}    \varepsilon_{it} \sim Skew-t( \zeta_{it}, \omega_{it},\lambda_{it},\nu)
   \end{array}
\end{equation}
\begin{equation}
        log(h_{it}) = \phi_{h,i}log(h_{it-1}) + \eta_{it} \hspace{2.5cm} \eta_{it} \sim \mathcal{N}(0, \sigma^2_{i,\eta}) \hspace{0.2cm} \text{\textit{i = gdpgrowth, NFCI}}
\end{equation}
\begin{equation}\label{gdpnfciskew}
\lambda_{gdpgrowth,t} =  \phi_{\lambda,1}\lambda_{gdpgrowth,t-1} + \beta_{1} NFCI_{t-1} + \xi_{gdpgrowth,t}   \hspace{1.3cm}  \xi_{i,t} \sim N(0, \sigma^2_{\xi,i})  
\end{equation}
\begin{equation}
\lambda_{NFCI,t} = \phi_{\lambda,2}\lambda_{NFCI,t-1} + \xi_{NFCI,t}     \hspace{6cm}  \xi_{i,t} \sim N(0, \sigma^2_{\xi,i}) 
\end{equation}

%\begin{equation}\label{gdpnfciskew}
% \lambda_{i,t} = \beta_0 + \lambda_{i,t-1} + \beta_{1} NFCI_{t-1} + \xi_{i,t}   \hspace{2cm}  \xi_{i,t} \sim N(0, \sigma^2_{\xi,i})  \hspace{0.2cm} \text{\textit{i = gdpgrowth}}
% \end{equation}
%
%\begin{equation}
% \lambda_{i,t} = \lambda_{i,t-1} + \xi_{i,t}     \hspace{4cm}  \xi_{i,t} \sim N(0, \sigma^2_{\xi,i})  \hspace{0.2cm} \text{\textit{i = NFCI}}
% \end{equation}
% 
In this VAR, the dynamic relationship between GDP growth and financial conditions in modelled jointly. In particular, in this specification, due to the triangular structure of $\boldsymbol{A}^{-1}$ shocks to GDP growth contemporaneously affect the financial markets, while shocks to NFCI do not affect GDP growth within the quarter. To understand whether the models perform well in forecasting downside risk, in what follows I will compare the forecast from the TVSSV models to the forecasts from the quantile regression based method of ABG. Their approach is based on a two step procedure where in the first step they use predictive quantile regression to estimate the quantiles of the conditional distribution:
\begin{equation}\label{QR}
    \hat{Q}_{gdpgrowth_{t+h}|\mathcal{I}_t}(\tau) = \boldsymbol{\hat{\beta}}^{\tau}X_{t}  \hspace{2cm} for \hspace{0.2cm} \tau = 0.05, \ldots, 0.95\ 
\end{equation}

Then, in the second step, the estimated quantiles are interpolated using a flexible \textit{Skew-t} distribution, so as to obtain a complete predictive density for GDP growth. We specify equation (\ref{QR}) collecting two lags of GDP growth and one lag of NFCI in the vector $X_{t}$, so as to capture changes in the future GDP growth distribution as a function of current financial and economic conditions. 

\subsection{Results}
\begin{figure}[H]
     \centering
     \caption{Posterior estimate of $\beta_1$}
     \includegraphics[scale = 0.25]{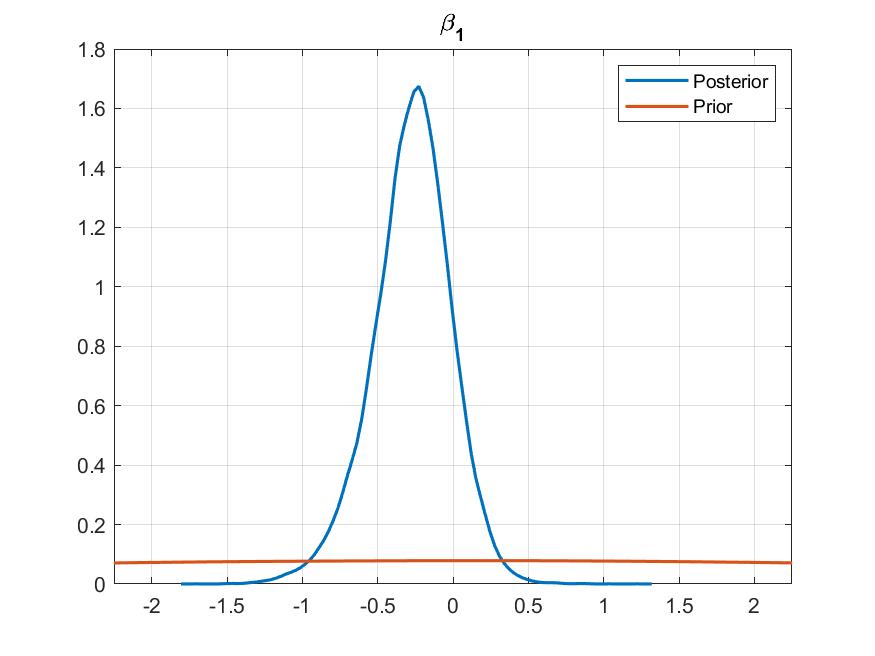}
    \floatfoot{\scriptsize Note:  The figure shows the estimated posterior distribution of the coefficient on the $NFCI_{t-1}$  in the equation of the shape parameter in the TVSSV model with \textit{Skew-t} shocks. }
     \label{fig:beta1}
 \end{figure}

\begin{figure}[H]
    \centering
    \caption{One quarter ahead Growth at Risk (GaR) and Expected Shortfall (1995Q1-2019Q4)}
    \includegraphics[scale = 0.5]{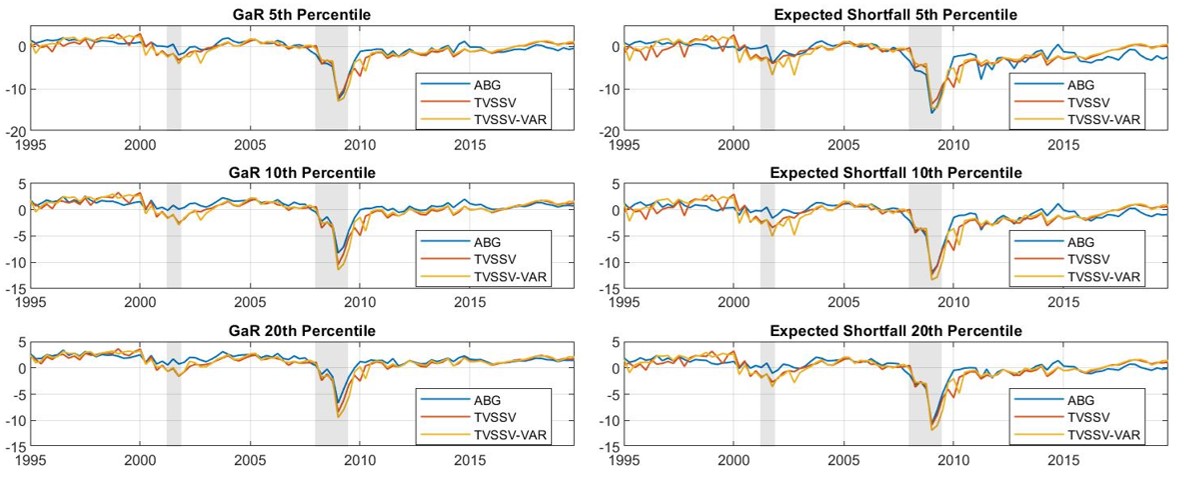}
    \floatfoot{\scriptsize Note: The figure shows the estimated $5^{th}$, $10^{th}$, $20^{th}$ percentiles of the one quarter ahead GDP growth predictive distribution (left panel) and the $5^{th}$, $10^{th}$, $20^{th}$ one quarter ahead expected shortfall (right panel). In blue estimates from the two step quantile regression based method by ABG, in red from the TVSSV univariate model with \textit{Skew-t} shocks and  in yellow the estimates from the TVSSV VAR model.}
    \label{fig:ES_GAR}
\end{figure}

\begin{figure}[H]
    \centering
    \caption{One quarter ahead recession probability (1995Q1-2019Q4)}
    \includegraphics[scale=0.35]{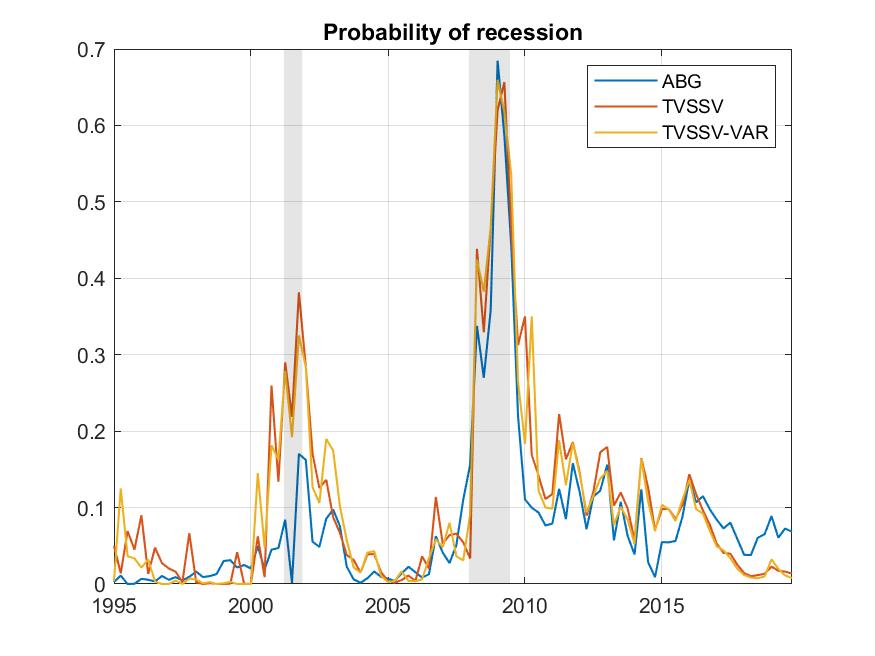}
    \floatfoot{\scriptsize Note:  The figure shows the estimated one quarter ahead recession probabilities. In blue estimates from the two step quantile regression based method by ABG, in red from the TVSSV univariate model with \textit{Skew-t} shocks and  in yellow the estimates from the TVSSV VAR model. }
    \label{fig:rec_prob}
\end{figure}

This section presents the results from the estimates of both the univariate TVSSV models and the VAR TVSSV models with \textit{Skew-Normal} and \textit{Skew-t}  shocks. The estimation sample starts in 1971Q1 and the forecasting exercise covers the period 1995Q1 - 2019Q4. Fig. \ref{fig:beta1} presents the estimated posterior distribution for the coefficient $\beta_1$ from the univariate time varying skewness stochastic volatility model. This is the coefficient that in the state equation of the skewness parameter (\ref{gdpnfciskew}) summarizes how the shape of the conditional distribution of GDP growth changes as a function of financial conditions in the previous quarter. As shown in Fig. \ref{fig:beta1},  tighter financial conditions (increases in the $NFCI$) are on average associated to a decrease in the skewness of current GDP growth (the posterior mean estimate is $\hat{\beta_1} = - 0.26)$. Hence, equation (\ref{skewness_nfci}),  captures  the main finding of ABG, which is that deteriorating financial condition are associated to movements in the lower quantiles of future GDP growth distribution. 
Ascertained that the model is able to capture the same asymmetric effect of financial conditions on the future GDP growth distribution found in ABG, it is important to understand what is the potential of the model to assess and predict risk out of sample. Fig. \ref{fig:ES_GAR} shows the out-of-sample forecasts of Growth at Risk and Expected Shortfall for the $5^{th}$, $10^{th}$ and $20^{th}$ percentiles while Fig. \ref{fig:rec_prob} shows the one quarter ahead estimated recession probability. I report the results from the stochastic volatility stochastic skewness model with \textit{Skew-t} shocks,  since the results from the model with \textit{Skew-Normal} shocks do not differ qualitatively. The figure shows that during the Financial Crisis our parametric models predicts as much downside risk to GDP growth as the quantile regression method of ABG. As shown in Fig. \ref{fig:rec_prob} both the TVSSV and the TVSSV-VAR models, assign higher probability of recession to the mild contraction of the U.S. economy following the dotcom bubble in 2000s with respect to the two step method based on quantile regression. 

In terms of forecast accuracy,  Table \ref{tab:oos} compares the forecasts from our parametric approaches to the forecasts from the method of ABG. The first two columns report the results for the average Log Scores and the average Cumulative Ranked Probability Scores (CRPS), since these two measures are the most commonly used to evaluate the relative density forecast accuracy of different models. \footnote{Defining $y$ the realization of the series to predict, $f(.)$  the density forecast and $F(.)$ corresponding the cumulative distribution, Logscores and CRPS are respectively defined as:

\begin{equation}
Logscores(f,y) =  - log(f(y))
\end{equation}

\begin{equation}
CRPS(f,y) = \int_{-\infty}^{\infty}PS(F(z),\mathbbm{1}\{ y \leq z\}) dz = \int_{0}^1 QS_{\alpha}(F^{-1}(\alpha),y)d\alpha
\end{equation}
where $PS(F(z),\mathbbm{1}\{ y \leq z \}) = (F(z) - \mathbbm{1}\{ y \leq z\})^2$ is the Brier probability score and $QS_{\alpha}(F^{-1}(\alpha), y) = 2 (\mathbbm{1}\{ y \leq F^{-1}(\alpha)\} - \alpha)(F^{-1}(\alpha) - y)$ is the Quantile Score.} 
Looking at average Logscores, the first column reports the difference between the forecasts from two step procedure of ABG and the forecasts from the time varying skewness stochastic volatility models (values greater than zero are associated to more accurate density forecast w.r.t ABG). According to the average Log-scores, our parametric models provide more accurate one quarter ahead density forecasts with respect to ABG. In parenthesis I report the \textit{p-values} from the Diebold and Mariano test \citep{DM} of equal forecast accuracy and find that for the TVSSV with \textit{Skew-t} shocks I am able to reject the null hypothesis of equal forecast accuracy. 
For what concerns average CRPS, on the second column,  the table reports the ratio with respect to the model of ABG (values lower than 1 are associated to more accurate density forecast with respect to ABG). As you can notice, based on this metrics, the time varying skewness stochastic volatility models perform as good if not even better than the two step procedure based on quantile regression. However, in all the cases I am not able to reject the null of equal forecast accuracy. 

Since I aim to assess the ability of the model to correctly characterize downside risk predictions, on the third column I report the average Quantile Weighted CRPS introduced by \citet{TWCRS} \footnote{The Quantile Weighted CRPS are computed as:

\begin{equation}
twCRPS = \int_{-\infty}^{\infty}PS(F(z),\mathbbm{1}\{ y \leq z\})^2w(z)dz  = \int_{0}^1 QS_{\alpha}(F^{-1}(\alpha),y)v(\alpha) d\alpha
\end{equation}

where $v(\alpha) = (1 - \alpha)^2$ assigns higher weights to the lower quantiles of the distribution function.} and on the fourth, fifth and sixth column I report the average Quantile Scores for the $5^{th}$, $10^{th}$ and $20^{th}$ percentiles commonly associated with the tick loss function \citep{GIAKOMU}. Also in this case I report the ratio with respect to the two step approach based on quantile regression (values lower than 1 are associated to more accurate density forecast with respect to ABG) and the \textit{p-values} from the Diebold-Mariano test in parenthesis. As you can notice, in terms of the ability of the model to correctly characterize downside risk predictions, I find that the stochastic volatility models performs comparably if not even better than ABG. In particular for the TVSSV-VAR with \textit{Skew-t} shocks I am able to reject the null of equal forecast accuracy with respect to ABG. The time series with the CRPS and left Tail Weighted CRPS, can be found in the Appendix \ref{otherf} (Fig. \ref{fig:crps_TS}). As well, in the Appendix \ref{otherf} the histogram with the PITs (Fig. \ref{fig:PITs}) reveals that the forecasts from the TVSSV models, are  better-calibrated with respect to the forecasts from the two-step quantile regression based method. Summing up, TVSSV models are able to reproduce the main finding in ABG, namely that deteriorating financial conditions are associated to shifts of the lower quantiles of the future GDP growth distribution. At the same time TVSSV models perform comparably if not even better than quantile regression based methods for forecasting macroeconomic tail risk.

%In summary, our parametric models represent a competitive alternative to quantile regression based methods for modelling and forecasting downside risk. 

\begin{table}[H]\scriptsize
    \centering \caption{One quarter ahead out of sample forecasts (1995Q1-2019Q4)}
    \begin{tabular}{l|ccccccc|}
    \hline
        ~  & \textbf{Log scores} & \textbf{CRPS} & \textbf{TwL CRPS} & $\boldsymbol{QS 5^{th}}$ & $\boldsymbol{QS 10^{th}}$ & $\boldsymbol{QS 20^{th}}$ \\ \hline \hline
\textbf{ABG} & 2.4840   &      1.1943    &    0.3623    &  0.2503   & 0.3926  &   0.5842 \\
\hline
        \textbf{TVSSV \textit{Skew Normal}}       & 0.1946    & 0.9757    & 0.9777    & 1.0334    & 0.9858    & 0.9832 \\
                                 & (0.1788)    & (0.7380)    & (0.2282)    & (0.6027)    & (0.3838)    & (0.3278) \\ \hline
\textbf{TVSSV \textit{Skew-t}}            & \textbf{0.3530}   & 0.9609    & 0.9823    & 1.0334    & 0.9911    & 0.9882 \\
                                 & \textbf{(0.0276)}    & (0.8659)    & (0.3045)    & (0.4725)    & (0.4472)    & (0.3934) \\ \hline
\textbf{TVSSV VAR \textit{Skew Normal}}   & 0.1620    & 0.9805    & 0.9644    & 0.9678    & 0.9718    & 0.9666 \\
                                 & (0.2287)    & (0.6781)    & (0.1669)    & (0.4038)    & (0.3209)    & (0.2589) \\ \hline
\textbf{TVSSV VAR \textit{Skew-t}}            & 0.0662    & 0.9700    & \textbf{0.9610}    & 1.0119    & 0.9979    & 0.9633 \\
                                 & (0.1050)    & (0.5942)    & \textbf{(0.0366)}    & (0.5493)    & (0.4778)    & (0.1052) \\
    \hline\hline 
    \multicolumn{7}{l}{\footnotesize Note: For the average Logscores, the first row reports the values from the ABG method  while the} \\
    \multicolumn{7}{l}{\footnotesize other rows report the difference between the two step procedure and the time varying skewness}\\
    \multicolumn{7}{l}{\footnotesize stochastic volatility models. For the other metrics I report the ratio w.r.t the ABG method.} \\
    \multicolumn{7}{l}{\footnotesize   Inside the parenthesis \textit{p-values} from the one sided Diebold-Mariano w.r.t the two step method} \\
    \multicolumn{7}{l}{ \footnotesize of \citet{ABG}.   The bold character indicates rejection of equal forecast accuracy at 5\%.}  
    \end{tabular}
    \label{tab:oos}
\end{table}

\section{Time varying skewness in a medium scale VAR} \label{sec: medscale}

\begin{figure}[H]
    \centering
    \caption{Estimated volatilities}\label{vola}
    \includegraphics[scale = 0.5]{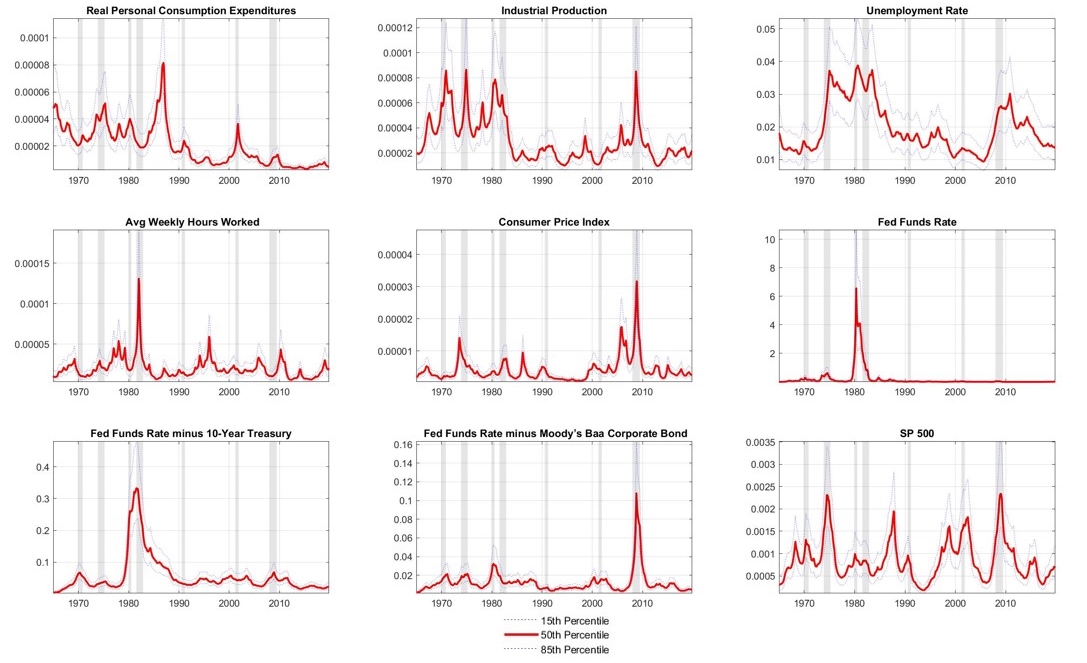}
    \floatfoot{\scriptsize Note: The figure shows the estimated volatilities  of the shocks in the TVSSV-VAR with \textit{Skew-t} shocks. In red the estimated median, in blue dashed lines the $85^{th}-15^{th}$ credible sets. }
\end{figure}

\begin{figure}[H]
    \centering
    \caption{Estimated shape parameters $\lambda_t$}\label{shape}
    \includegraphics[scale = 0.5]{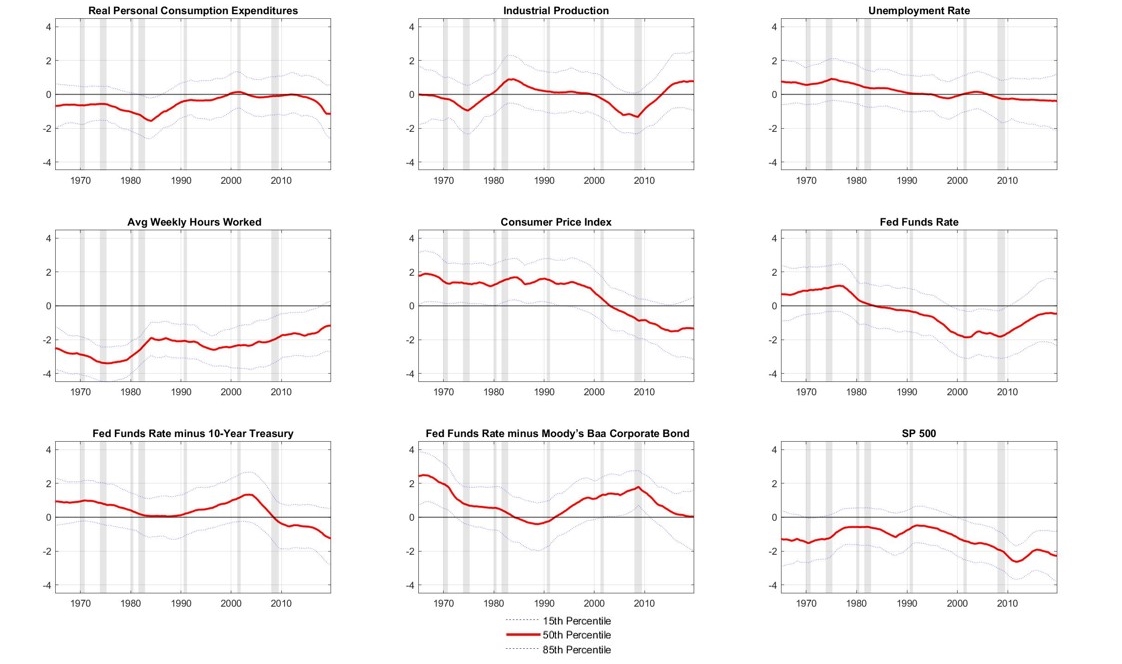}
    \floatfoot{\scriptsize Note:  The figure shows the estimated shape parameters of the shocks in the TVSSV-VAR with \textit{Skew-t} shocks. In red the estimated median, in blue dashed lines the $85^{th}-15^{th}$ credible sets. }
\end{figure}

One of the main advantages of the VAR model presented in Section \ref{sec:VAR} is that it allows to explicitly capture time varying conditional skewness of multiple time series. In this section I estimate a medium scale VAR model which includes macroeconomic and financial monthly time series and I investigate the time varying asymmetric behaviour of the shocks to the variables in the system.      I consider a VAR model with 8 variables being Real personal consumption expenditures, Industrial Production, Unemployment Rate, average Weekly Hours Worked, Consumer Price Index, Fed Funds Rate, the spread between 10-Year Treasury and the Fed Funds Rate, the spread between Moody’s Baa Corporate Bond and the Fed Funds Rate and the Standard and Poors Index. The variables are in monthly frequency and are taken from the  FRED-MD.\footnote{Table \ref{tab:trans} in the Appendix reports the variable transformation.}
%
%We start with  preliminary analysis on the residuals of a baseline Stochastic Volatility Model . In particular, I study the properties of the standardized residuals  $\bar{u}_{it} = \frac{u_{it}}{\sqrt{h}_{it}}$ by means of the QQ-Plots. As shown by Fig. ** in the Appendix, for most of the series the standardized residuals  exhibits  excess kurtosis and skewness. 
%
I present the results from the VAR with \textit{Skew-t} shocks.\footnote{For the VAR with \textit{Skew-Normal} the estimated path for the volatilities and shape parameters are almost the same.} I include 13 lags and assume a Minnesota prior structure for the variance covariance matrix of the regression coefficients. \footnote{See the Appendix for the details on the hyper-parameters of the Minnetota Prior. } %Conditioning on the previous lags and on the contemporaneous relationship between the variables in the system implied by the matrix $\boldsymbol{A}$, the distribution of the endogenous variables in the system is not symmetric when $\lambda_{i,t} \neq 0 $. In other words, when $\lambda_{i,t} \neq 0 $ the variables in the system are conditionally skewed. Hence, the shape parameter $\lambda_{i,t}$ can be interpreted as a persistent and exogenous measure of risk. \\
The estimation sample is January 1965 - December 2019. Fig. \ref{vola} shows the estimated volatilities while Fig. \ref{shape} shows the estimated shape parameters. The dotted line in blue are the $85^{th}-15^{th}$ credible sets while the red line is the estimated posterior median. 

It is interesting to notice that shocks to the CPI were on average positively skewed before the 2000s while became left skew for the rest of the sample that ends on 2019. This switch in the sign of the shape parameter indicates that conditionally on the past and on the contemporaneous realization of Real Personal Consumption Expenditures, Industrial Production, Unemployment Rate and average Weekly Hours Worked, the distribution of CPI was right skewed in the 1980s, becoming instead left skew from the 2000s.  In other words, risk switched from the upside  to the downside.   
As for the monetary policy shocks, in the 1980s large positive  hikes of the Fed Fund Rate were more frequent, while from the early 2000s large negative shocks to the Fed Fund Rate become more likely. 
Shocks to the average Weekly Hours Worked are skewed to the left over the entire sample, which means that negative large shocks have been systematically more frequent than positive large shocks. As well, shocks to the stock market (SP 500 index) are skewed to the left over the entire sample. This is in line with the large body of the financial econometrics literature that studies conditional skewness in asset returns \citep{harvey2000conditional}. 
As for the spread between 10-Year Treasury and the Fed Funds Rate and the spread between Moody’s Baa Corporate Bond and the Fed Funds Rate, for most of the sample both the shocks are skewed to the right meaning that the probability of large positive shocks has been greater than the probability of large negative shocks. This finding vanishes starting from 2009 and might be linked to the unconventional monetary policy following the Great Financial Crisis.

%\subsection{\textcolor{red}{Structural Scenario Analysis}}

%\textcolor{red}{In progress [...]}

\section{Conclusion}

In this paper I propose a fully parametric framework based on time varying skewness stochastic volatility models with \textit{Skew-Normal} and \textit{Skew-t} shocks for assessing and forecasting macroeconomic tail risk. First, I consider an extension of the univariate stochastic volatility model of \citet{jpr1993} that explicitly accounts for time varying skewness in the predictive distribution of the dependent variable. Then, I introduce a Bayesian VAR model with stochastic volatility and stochastic skewness to provide an explicit treatment of conditional skewness when modelling the dynamics of multiple time series. I compare the time varying skewness stochastic volatility models to the quantile regression method of \citet{ABG} to assess and predict tail risk to GDP growth. I find that the time varying skewness stochastic volatility models considered in this paper are able to reproduce the main findings of \citet{ABG}, that is the nonlinear and asymmetric effect of financial conditions on the future GDP growth distribution. The models predict as much risk as quantile regression  during the Financial crisis while provide slightly more accurate out of sample forecasts of downside risk over the entire sample. Finally, estimating a standard medium scale VAR model I find that time varying skewness is a relevant feature of macroeconomic and financial shocks. \\

\textbf{Future research} \hspace{0.2cm}  For future research, the VAR model considered in this paper could be used to study the probability of joint tail events and for constructing structural scenarios of ``at-risk'' measures. For example, it could be used to study and assess stagflation risk, or to analyze scenarios for inflation at-risk and labour-at risk under different monetary policy paths.  As a methodological extension, particularly interesting would be to consider an order invariant version of this model. 
%\textcolor{red}{We show that allowing for skewness is particularly use full to inform policymakers.... in Structural Scenario Analysis  In progress [...]}
\clearpage
\printbibliography

@article{COGLEY2005262,
title = {Drifts and volatilities: monetary policies and outcomes in the post WWII US},
journal = {Review of Economic Dynamics},
volume = {8},
number = {2},
pages = {262-302},
year = {2005},
note = {Monetary Policy and Learning},
issn = {1094-2025},
doi = {https://doi.org/10.1016/j.red.2004.10.009},
url = {https://www.sciencedirect.com/science/article/pii/S1094202505000049},
author = {Timothy Cogley and Thomas J. Sargent}
}

@article{azzalini1986,
  title={Further results on a class of distributions which includes the normal ones},
  author={Azzalini, Adelchi},
  journal={Statistica},
  volume={46},
  number={2},
  pages={199--208},
  year={1986}
}

@article{primiceri2005,
  title={Time varying structural vector autoregressions and monetary policy},
  author={Primiceri, Giorgio E},
  journal={The Review of Economic Studies},
  volume={72},
  number={3},
  pages={821--852},
  year={2005},
  publisher={Wiley-Blackwell}
}

@article{carriero2019large,
  title={Large Bayesian vector autoregressions with stochastic volatility and non-conjugate priors},
  author={Carriero, Andrea and Clark, Todd E and Marcellino, Massimiliano},
  journal={Journal of Econometrics},
  volume={212},
  number={1},
  pages={137--154},
  year={2019},
  publisher={Elsevier}
}

@article{lindsten2014particle,
  title={Particle Gibbs with ancestor sampling},
  author={Lindsten, Fredrik and Jordan, Michael I and Schon, Thomas B},
  journal={Journal of Machine Learning Research},
  volume={15},
  pages={2145--2184},
  year={2014},
  publisher={MICROTOME PUBL}
}

@article{andrieu2010particle,
  title={Particle markov chain monte carlo methods},
  author={Andrieu, Christophe and Doucet, Arnaud and Holenstein, Roman},
  journal={Journal of the Royal Statistical Society: Series B (Statistical Methodology)},
  volume={72},
  number={3},
  pages={269--342},
  year={2010},
  publisher={Wiley Online Library}
}

@article{A2015,
  title={Bayesian estimation of a skew-student-t stochastic volatility model},
  author={Abanto-Valle, CA and Lachos, VH and Dey, Dipak K},
  journal={Methodology and Computing in Applied Probability},
  volume={17},
  number={3},
  pages={721--738},
  year={2015},
  publisher={Springer}
}

@article{AC2003,
author = {Azzalini, Adelchi and Capitanio, Antonella},
title = {Distributions generated by perturbation of symmetry with emphasis on a multivariate skew t-distribution},
journal = {Journal of the Royal Statistical Society: Series B (Statistical Methodology)},
volume = {65},
number = {2},
pages = {367-389},
doi = {https://doi.org/10.1111/1467-9868.00391},
url = {https://rss.onlinelibrary.wiley.com/doi/abs/10.1111/1467-9868.00391},
eprint = {https://rss.onlinelibrary.wiley.com/doi/pdf/10.1111/1467-9868.00391},
year = {2003}
}

@article{jpr1993,
title = {Bayesian Analysis of Stochastic Volatility Models},
author = {Jacquier, Eric and Polson, Nicholas G and Rossi, Peter},
year = {1994},
journal = {Journal of Business \& Economic Statistics},
volume = {12},
number = {4},
pages = {371-89},
url = {https://EconPapers.repec.org/RePEc:bes:jnlbes:v:12:y:1994:i:4:p:371-89}
}

@ARTICLE{JPR2004,
title = {Bayesian analysis of stochastic volatility models with fat-tails and correlated errors},
author = {Jacquier, Eric and Polson, and Rossi, Peter},
year = {2004},
journal = {Journal of Econometrics},
volume = {122},
number = {1},
pages = {185-212},
url = {https://EconPapers.repec.org/RePEc:eee:econom:v:122:y:2004:i:1:p:185-212}
}

@article{CappuccioLubianRaggi2004,
author = {Nunzio Cappuccio and Diego Lubian and Davide Raggi},
doi = {doi:10.2202/1558-3708.1211},
url = {https://doi.org/10.2202/1558-3708.1211},
title = {MCMC Bayesian Estimation of a Skew-GED Stochastic Volatility Model},
journal = {Studies in Nonlinear Dynamics \& Econometrics},
number = {2},
volume = {8},
year = {2004}
}

@article{ISERINGHAUSEN2020,
title = {The time-varying asymmetry of exchange rate returns: A stochastic volatility – stochastic skewness model},
journal = {Journal of Empirical Finance},
volume = {58},
pages = {275-292},
year = {2020},
issn = {0927-5398},
doi = {https://doi.org/10.1016/j.jempfin.2020.06.008},
url = {https://www.sciencedirect.com/science/article/pii/S0927539820300396},
author = {Martin Iseringhausen}
}

@TechReport{petrelladellemonache2021,
  author={Delle Monache, Davide and De Polis, Andrea and Petrella, Ivan},
  title={{Modeling and forecasting macroeconomic downside risk}},
  year=2021,
  month=Mar,
  institution={Bank of Italy, Economic Research and International Relations Area},
  type={Temi di discussione (Economic working papers)},
  url={https://ideas.repec.org/p/bdi/wptemi/td_1324_21.html},
  number={1324},
  doi={},
}

@TechReport{loria2019,
  author={J. David López-Salido and Francesca Loria},
  title={{Inflation at Risk}},
  year=2020,
  month=Feb,
  institution={Board of Governors of the Federal Reserve System (U.S.)},
  type={Finance and Economics Discussion Series},
  url={https://ideas.repec.org/p/fip/fedgfe/2020-13.html},
  number={2020-013},
  doi={10.17016/FEDS.2020.013},
}

@article{BROWNLEES2021312,
title = {Backtesting global Growth-at-Risk},
journal = {Journal of Monetary Economics},
volume = {118},
pages = {312-330},
year = {2021},
issn = {0304-3932},
doi = {https://doi.org/10.1016/j.jmoneco.2020.11.003},
url = {https://www.sciencedirect.com/science/article/pii/S0304393220301288},
author = {Christian Brownlees and André B.M. Souza}
}

@article{ABG,
Author = {Adrian, Tobias and Boyarchenko, Nina and Giannone, Domenico},
Title = {Vulnerable Growth},
Journal = {American Economic Review},
Volume = {109},
Number = {4},
Year = {2019},
Month = {4},
Pages = {1263-89},
DOI = {10.1257/aer.20161923},
URL = {https://www.aeaweb.org/articles?id=10.1257/aer.20161923}}

@article{GIGLIO2016457,
title = {Systemic risk and the macroeconomy: An empirical evaluation},
journal = {Journal of Financial Economics},
volume = {119},
number = {3},
pages = {457-471},
year = {2016},
issn = {0304-405X},
doi = {https://doi.org/10.1016/j.jfineco.2016.01.010},
url = {https://www.sciencedirect.com/science/article/pii/S0304405X16000143},
author = {Stefano Giglio and Bryan Kelly and Seth Pruitt}
}

@TechReport{CCM2020BVAR,
  author={Andrea Carriero and Todd E. Clark and Massimiliano Marcellino},
  title={{Capturing Macroeconomic Tail Risks with Bayesian Vector Autoregressions}},
  year=2020,
  month=Jan,
  institution={Federal Reserve Bank of Cleveland},
  type={Working Papers},
  url={https://ideas.repec.org/p/fip/fedcwq/87375.html},
  number={20-02R},
  doi={10.26509/frbc-wp-202002r},
}

@article{GELOS2022103555,
title = {Capital flows at risk: Taming the ebbs and flows},
journal = {Journal of International Economics},
volume = {134},
pages = {103555},
year = {2022},
issn = {0022-1996},
doi = {https://doi.org/10.1016/j.jinteco.2021.103555},
url = {https://www.sciencedirect.com/science/article/pii/S0022199621001355},
author = {Gaston Gelos and Lucyna Gornicka and Robin Koepke and Ratna Sahay and Silvia Sgherri}
}

@article{kiley2018unemployment,
  title={Unemployment risk},
  author={Kiley, Michael T},
  journal={Journal of Money, Credit and Banking},
  year={2018},
  publisher={Wiley Online Library}
}

@article{TWCRS,
 ISSN = {07350015},
 URL = {http://www.jstor.org/stable/23243806},
 author = {Tilmann Gneiting and Roopesh Ranjan},
 journal = {Journal of Business \& Economic Statistics},
 number = {3},
 pages = {411--422},
 publisher = {American Statistical Association},
 title = {Comparing Density Forecasts Using Threshold-and Quantile-Weighted Scoring Rules},
 volume = {29},
 year = {2011}
}

@article{GIAKOMU,
 ISSN = {07350015},
 URL = {http://www.jstor.org/stable/27638838},
 author = {Raffaella Giacomini and Ivana Komunjer},
 journal = {Journal of Business \& Economic Statistics},
 number = {4},
 pages = {416--431},
 publisher = {[American Statistical Association, Taylor \& Francis, Ltd.]},
 title = {Evaluation and Combination of Conditional Quantile Forecasts},
 volume = {23},
 year = {2005}
}

@article{CLARKRAV2015,
author = {Clark, Todd E. and Ravazzolo, Francesco},
title = {Macroeconomic Forecasting Performance under Alternative Specifications of Time-Varying Volatility},
journal = {Journal of Applied Econometrics},
volume = {30},
number = {4},
pages = {551-575},
doi = {https://doi.org/10.1002/jae.2379},
url = {https://onlinelibrary.wiley.com/doi/abs/10.1002/jae.2379},
eprint = {https://onlinelibrary.wiley.com/doi/pdf/10.1002/jae.2379},
year = {2015}
}

@article{KARLSSON,
title = {Vector autoregression models with skewness and heavy tails},
journal = {Journal of Economic Dynamics and Control},
volume = {146},
pages = {104580},
year = {2023},
issn = {0165-1889},
doi = {https://doi.org/10.1016/j.jedc.2022.104580},
url = {https://www.sciencedirect.com/science/article/pii/S0165188922002834},
author = {Karlsson, Sune  and Mazur, Stepan and Nguyen,Hoang}
}

@ARTICLE{DM,
title = {Comparing Predictive Accuracy},
author = {Diebold, Francis and Mariano, Roberto},
year = {1995},
journal = {Journal of Business \& Economic Statistics},
volume = {13},
number = {3},
pages = {253-63},
url = {https://EconPapers.repec.org/RePEc:bes:jnlbes:v:13:y:1995:i:3:p:253-63}
}

@article{harvey2000conditional,
  title={Conditional skewness in asset pricing tests},
  author={Harvey, Campbell R and Siddique, Akhtar},
  journal={The Journal of finance},
  volume={55},
  number={3},
  pages={1263--1295},
  year={2000},
  publisher={Wiley Online Library}
}

@article{SW2012,
Author = {Stock, James H. and Watson, Mark W.},
Title = {Vector Autoregressions},
Journal = {Journal of Economic Perspectives},
Volume = {15},
Number = {4},
Year = {2001},
Month = {12},
Pages = {101-115},
DOI = {10.1257/jep.15.4.101},
URL = {https://www.aeaweb.org/articles?id=10.1257/jep.15.4.101}}

@article{corrCARRIERO2022,
title = {Corrigendum to “Large Bayesian vector autoregressions with stochastic volatility and non-conjugate priors” [J. Econometrics 212 (1) (2019) 137–154]},
journal = {Journal of Econometrics},
volume = {227},
number = {2},
pages = {506-512},
year = {2022},
issn = {0304-4076},
doi = {https://doi.org/10.1016/j.jeconom.2021.11.010},
url = {https://www.sciencedirect.com/science/article/pii/S0304407621002773},
author = {Andrea Carriero and Joshua Chan and Todd E. Clark and Massimiliano Marcellino}
}

@TechReport{km, author={Kilian, Lutz and Manganelli, Simone}, title={{The central bank as a risk manager: quantifying and forecasting inflation risks}}, year=2003, month=Apr, institution={European Central Bank}, type={Working Paper Series}, url={https://ideas.repec.org/p/ecb/ecbwps/2003226.html}, number={226}, doi={}, }

@article{wolf2021estimating,
  title={Estimating growth at risk with skewed stochastic volatility models},
  author={Wolf, Elias},
  journal={Available at SSRN 4030094},
  year={2021}
}

@incollection{montes2022skewed,
  title={Skewed SVARS: Tracking the structural sources of macroeconomic tail risks},
  author={Montes-Gald{\'o}n, Carlos and Ortega, Eva},
  booktitle={Essays in Honour of Fabio Canova},
  volume={44},
  pages={177--210},
  year={2022},
  publisher={Emerald Publishing Limited}
}

@article{iseringhausen2021time,
  title={A time-varying skewness model for Growth-at-Risk},
  author={Iseringhausen, Martin},
  year={2021},
  publisher={European Stability Mechanism Working Paper}
}

@TechReport{shin2021,
  author={Jonas E. Arias and Juan F. Rubio-Ramirez and Minchul Shin},
  title={{Macroeconomic Forecasting and Variable Ordering in Multivariate Stochastic Volatility Models}},
  year=2021,
  month=Jun,
  institution={Federal Reserve Bank of Philadelphia},
  type={Working Papers},
  url={https://ideas.repec.org/p/fip/fedpwp/92355.html},
  number={21-21},
  doi={10.21799/frbp.wp.2021.21},
}

@article{chan2021large,
  title={Large order-invariant Bayesian VARs with stochastic volatility},
  author={Chan, Joshua CC and Koop, Gary and Yu, Xuewen},
  journal={arXiv preprint arXiv:2111.07225},
  year={2021}
}

@article{sims1980macroeconomics,
  title={Macroeconomics and reality},
  author={Sims, Christopher A},
  journal={Econometrica: journal of the Econometric Society},
  pages={1--48},
  year={1980},
  publisher={JSTOR}
}

@article{ccm2015,
author = {Carriero, Andrea and Clark, Todd E. and Marcellino, Massimiliano},
title = {Realtime nowcasting with a Bayesian mixed frequency model with stochastic volatility},
journal = {Journal of the Royal Statistical Society: Series A (Statistics in Society)},
volume = {178},
number = {4},
pages = {837-862},
doi = {https://doi.org/10.1111/rssa.12092},
url = {https://rss.onlinelibrary.wiley.com/doi/abs/10.1111/rssa.12092},
eprint = {https://rss.onlinelibrary.wiley.com/doi/pdf/10.1111/rssa.12092},
year = {2015}
}
\clearpage
\appendix
\section{Appendix}
\subsection{Skew Normal and Skew-t: distributions and parameterization}\label{sec:appendixA}

The \textit{Skew Normal} \citep{azzalini1986} distribution is: 

\[
\begin{aligned}
& p(\varepsilon_t|\zeta, \omega^2, \lambda) = \frac{2}{\omega} \phi\left(\frac{\varepsilon_t - \zeta}{\omega}\right) \Phi\left( \lambda \left(\frac{ \varepsilon_t - \zeta}{\omega} \right) \right)\\
\end{aligned}
\]
where $\phi(.)$ and $\Phi(.)$ are respectively the pdf and cdf of the standard Normal. In general, $\varepsilon_t \sim Skew-Normal( \zeta, \omega^2, \lambda) $  has the following stochastic representation: 
\begin{equation}\label{rapsn}
    \varepsilon_{t} =  \zeta + \delta \omega v_{t} + \sqrt{(1-\delta^2)}\omega z_{t}
\end{equation}
where: 

$v_{t} \stackrel{i.i.d}{\sim} \text{Truncated Normal}_{[0,\infty)}(0,1) $

$z_{t} \stackrel{i.i.d}{\sim} \mathcal{N}(0,1)$      

$\delta = \frac{\lambda}{\sqrt{1 + \lambda^2}}$, with $-1<\delta<1$.  \\

The mean and the variance of $\varepsilon_{t}$ are given by: 
\begin{equation}
    \mathbb{E}[\varepsilon_t]  = \zeta + \omega \delta \sqrt{\frac{2}{\pi}}
\end{equation}   
\begin{equation}
var(\varepsilon_t)  = \omega^2\left( 1 - \frac{2\delta^2}{\pi} \right)   
\end{equation}
Assuming $\E[\varepsilon_t] = 0$ and $var(\varepsilon_t) = 1$ leads to the following constraints on the location and scale parameters: $\zeta = - \omega \delta \sqrt{\frac{2}{\pi}}$
and $\omega^2 = \left( 1 - \frac{2\delta^2}{\pi}\right)^{-1}$. Once we impose these constraints on the location and scale parameters, with  $\lambda = 0$ the distribution collapses to the Standard Normal. \\

The \textit{Skew-t} distribution \citep{AC2003} is: 

\begin{equation}
    p(\varepsilon_t|\zeta, \omega^2, \lambda, \nu) = \frac{2}{\omega} t_\nu\left(\frac{\varepsilon_t - \zeta}{\omega}\right) T_{\nu + 1}\left( \lambda \left(\frac{ \varepsilon_t - \zeta}{\omega} \right) \sqrt{\frac{\nu +1}{\nu \left( \frac{ \varepsilon_t - \zeta }{\omega} \right)^2 }}   \right)\\
\end{equation}

where $t(.)$ and $T(.)$ are respectively the pdf and cdf of the Student-t with $\nu$ degrees of freedom.  $\varepsilon_{t} \sim \text{\textit{Skew-t}}(\zeta,\omega^2,\lambda, \nu)$ has the following stochastic representation: 

\begin{equation}
    \varepsilon_{t} =  \zeta + \delta\omega o_{t}^{-0.5}v_{t} + \sqrt{(1-\delta^2)}\omega o_{t}^{-0.5} z_{t}
\end{equation}

where: 

$v_{t} \stackrel{i.i.d}{\sim} \text{Truncated Normal}_{[0,\infty)}(0,1) $

$z_{t} \stackrel{i.i.d}{\sim} \mathcal{N}(0,1)$                        

$o_{t} \stackrel{i.i.d}{\sim} \mathcal{G}(\frac{\nu}{2},\frac{\nu}{2})  $ 

$\delta = \frac{\lambda}{\sqrt{1 + \lambda^2}}$, with $-1<\delta<1$.  \\

The mean and the variance of $\varepsilon_{t}$ are given by: 
\begin{equation}
   \mathbb{E}[\varepsilon_t]  =  \omega \delta k_1\sqrt{\frac{2}{\pi}} 
\end{equation}
\begin{equation}
    var(\varepsilon_t)  = \omega^2 \left(  k_2 - \frac{2}{\pi}k_1^2\delta^2 \right)
\end{equation}

with $k_1 = \sqrt{\frac{\nu}{2}}\frac{\Gamma(\frac{\nu-1}{2})}{\Gamma(\frac{\nu}{2})}$,  $k_2 = \frac{\nu}{\nu - 2}$. \\

Assuming $ \E[\varepsilon_t] = 0$ and $var(\varepsilon_t) = 1$ leads to the following constraints on the location and scale parameters:  $\zeta = - \omega \delta k_1\sqrt{\frac{2}{\pi}}$ and $\omega^2 = \left(  k_2 - \frac{2}{\pi}k_1^2\delta^2 \right)^{-1}$. Once we impose these constraints on the location and scale parameters with $\lambda= 0$ the distribution collapses to a \textit{Student-t} distribution properly re-scaled to have unit variance (and zero mean).

%\begin{figure}[H] 
%    \centering
%    \caption{The \textit{Skew Normal} and \textit{Skew-t} distribution}
%    \includegraphics[scale = 0.35]{skewness.jpg}
%    \floatfoot{\tiny Note:  The figure shows the \textit{Standard Normal}, \textit{Skew-Normal}, \textit{Student-t} and \textit{Skew-t}}
 %   \label{fig:skewn}
%\end{figure}

\subsection{Full conditional posterior distributions}\label{sec:appendixB}

\subsubsection{Univariate time varying skewness stochastic volatility model: skew normal shocks}

The full conditional distribution of $\{ v_{t}\}_{t=1}^T$ is given by: 
        
\begin{equation}\scriptsize
 p(v_t| .)   \propto exp \left[ -\frac{1}{2} \left( \frac{1}{(1-\delta_t^2)}v_t^2 - \frac{2\delta_t h_t^{-0.5}}{\omega_t(1-\delta_t^2)}(y_t - \boldsymbol{x_{t}}\boldsymbol{\pi} - \zeta_t\sqrt{h_t})v_t\right) \right] \mathbb{I}( 0  \leq v_t < \infty)
\end{equation}
this is a \textit{Truncated Normal}$ \left( \frac{\delta_t h_t^{-0.5}[y_t - \boldsymbol{x_{t}}\boldsymbol{\pi}] - \delta_t \zeta_t}{\omega_t}, 1-\delta_t^2\right)_{[0,\infty)}$ \\

The full conditional distribution of $\boldsymbol{\pi}$ is Normal: 
\begin{equation}
f(\boldsymbol{\pi}| .) \sim N (\overline{\mu}_{\pi} ,  \overline{\Sigma}_{\pi} )  \\
\end{equation}
%
%\begin{equation}
%\scriptsize
%f(\pi| .) &\propto exp \left[ - \frac{1}{2}  \sum_{t=1}^T  \frac{(\tilde{y}_t - x_{t}\beta)^2}{\sigma^2_t} \right] |\Sigma_{\beta}|^{-1} exp \left[ -\frac{1}{2} (\beta - \underline{\mu_{\beta}})'\underline{\Sigma_{\beta}^{-1}}(\beta - \underline{\mu_{\beta}}) \right]  \\
%\end{equation}
%
\begin{equation*}
\begin{array}{l}
 \overline{\mu}_{\pi} =  \overline{\Sigma}_{\pi} \left(\sum_{t=1}^T \frac{1}{\sigma^2_t} \boldsymbol{x_{t}}'\tilde{y}_t  + \underline{\Sigma_{\pi}}^{-1} \underline{\mu_{\pi}}\right)  \\
 \overline{\Sigma}_{\pi}^{-1} = \underline{\Sigma_{\pi}}^{-1} + \sum_{t=1}^T \frac{1}{\sigma^2_t} \boldsymbol{x_{t}}'\boldsymbol{x_{t}}   \\
\end{array}
\end{equation*}
where 

$ \tilde{y}_t \equiv y_t - \sqrt{h_t}\zeta_t - \sqrt{h_t}\omega_t \delta_t v_t$

$  \sigma^2_t \equiv h_t \omega^2_t (1-\delta_t^2)$
\newline\noindent
while  $\underline{\mu_{\pi}}$ and $\underline{\Sigma_{\pi}}$ are the prior mean and variance covariance matrix.\\

The full conditional distribution of $\phi_h$ is a \textit{Normal} :

\begin{equation}
    f(\phi_h|.) \sim \mathcal{N}( \bar{\mu}_{\phi_h} , \bar{\sigma}^2_{\phi_h} )
\end{equation}
\begin{equation}
\bar{\sigma}^2_{\phi_h} =  \left( \sum_{t = 1}^T\frac{log(h_{t-1})^2}{\sigma^2_{\eta}} + \frac{1}{\underline{\sigma^2_{\phi_h}}} \right)^{-1} 
\end{equation}

\begin{equation}
\bar{\mu}_{\phi_h} = \bar{\sigma}^2_{\phi_h}\left( \frac{\underline{\mu}_{\phi_h}}{\underline{\sigma^2_{\phi_h}}}  + \sum_{t = 1}^T\frac{log(h_{t-1})log(h_t)}{\sigma^2_{\eta}}\right)
\end{equation}

where $\underline{\mu}_{\phi_h}$ and $\underline{\sigma}^2_{\phi_h}$ are prior mean and variance.

The full conditional distribution of $\sigma^2_{\eta}$ is an \textit{Inverse Gamma} :

\begin{equation}
\scriptsize
  p(\sigma^2_{\eta}| .) \propto \left(\frac{1}{\sigma_{\eta}^2} \right)^{\frac{T}{2}} exp \left[ \sum_{t=1}^T-\frac{1}{2\sigma_{\eta,i}^2}(ln(h_{t}) - \phi_hln(h_{t-1}))^2 \right]   exp \left[-\frac{s_{\sigma_{\eta}^2}}{\sigma_{\eta}^2}\right] \sigma_{\eta}^{2^{-\nu_{\sigma_{\eta}^2}} - 1}   
\end{equation}

where $s_{\sigma_{\eta}^2}$ and $\nu_{\sigma_{\eta}}^2$ are the hyper-parameters of the Inverse Gamma prior.

The full conditional distribution of $\phi_{\lambda}$ is a \textit{Normal} :

\begin{equation}
    f(\phi_{\lambda}|.) \sim \mathcal{N}( \bar{\mu}_{\phi_{\lambda}} , \bar{\sigma}^2_{\phi_{\lambda}} )
\end{equation}
\begin{equation}
\bar{\sigma}^2_{\phi_{\lambda}} =  \left( \sum_{t = 1}^T\frac{\lambda_{t-1}^2}{\sigma^2_{\xi}} + \frac{1}{\underline{\sigma^2_{\phi_{\lambda}}}} \right)^{-1} 
\end{equation}

\begin{equation}
\bar{\mu}_{\phi_{\lambda}} = \bar{\sigma}^2_{\phi_{\lambda}}\left( \frac{\underline{\mu}_{\phi_{\lambda}}}{\underline{\sigma^2_{\phi_{\lambda}}}}  + \sum_{t = 1}^T\frac{{\lambda}_{t-1}{\lambda_t}}{\sigma^2_{\xi}}\right)
\end{equation}

where $\underline{\mu}_{\phi{\lambda}}$ and $\underline{\sigma}^2_{\phi_{\lambda}}$ are prior mean and variance.

The full conditional distribution of $\sigma^2_{\xi}$ is an \textit{Inverse Gamma}:

\begin{equation}
\scriptsize
 p(\sigma^2_{\xi}| .) \propto \left(\frac{1}{\sigma^2_{\xi}} \right)^{\frac{T}{2}}  exp \left[ \sum_{t=1}^T-\frac{1}{2\sigma^2_{\xi}}(\lambda_{it} - \phi_{\lambda}\lambda_{it-1})^2 \right] exp \left[-\frac{s_{\sigma_{\xi}^2}}{\sigma_{\xi}^2}\right] \sigma_{\xi}^{2^-{\nu_{\sigma_{\xi}^2}} - 1}    
\end{equation}

where $s_{\sigma_{\xi}^2}$ and $\nu_{\sigma_{\xi}^2}$ are the hyper-parameters of the Inverse Gamma prior.

The full conditional distribution of $h_0$ is $\mathcal{N} \left(\bar{\mu}_{h 0}, \bar{\sigma}_{h0}\right) $
\begin{equation}
\bar{\mu}_{h0} = \bar{\sigma}_{h0}\left(\frac{\mu_{h0}}{\sigma^2_{h0}} + \frac{\frac{log(h_1)}{\phi_h}}{\frac{\sigma^2_{\eta}}{\phi_{h}^2}}\right)  
\end{equation}

\begin{equation}
\bar{\sigma}_{h0} = \frac{\frac{\sigma^2_{h_0}\sigma^2_{\xi}}{\phi^2_{h}}}{\sigma^2_{h0}  + \frac{\sigma^2_{\eta}}{\phi^2_{h}}}  
\end{equation}

where $\mu_{\lambda_0}$ and $\sigma^2_{\lambda_0}$ are the prior mean and variance.

The full conditional distribution of $\boldsymbol{h}$ is given by: 
\begin{equation}
\scriptsize
 p(\boldsymbol{h} | .)  = \prod_{t=1}^T p(h_t | h_{t-1}, h_{t+1},.) \\
\end{equation}

\begin{equation}
\scriptsize
 p(h_t|.)  \propto h_t ^{-1,5}  exp \left[\left(-\frac{1}{2}\left(\frac{y_{t} - \boldsymbol{x_{t}} \boldsymbol{\pi} - \sqrt{h_{t}}\zeta_{t}  - \sqrt{h_{t} }\omega_{t}  \delta_{t}  v_{t}}{\sqrt{h_{t} } \omega_{t}  (1-\delta_{t} ^2)^{0.5}}\right)^2  -\frac{1}{2} \frac{(ln h_t - \mu_{h_t})^2}{\sigma_{h_t}^2}       \right)  \right]
\end{equation}

\begin{equation}\label{propn}
\begin{array}{l}
\mu_{h_t} = \frac{\phi_h}{\phi_h^2+1} \left( ln h_{t+1} + ln h_{t-1} \right) \\
\sigma_{h}^2 = \frac{\sigma^2_{\eta}}{\phi_h^2+1}
\end{array}
\end{equation}

The full conditional distribution of $\lambda_0$ is $\mathcal{N} \left(\bar{\mu}_{\lambda 0}, \bar{\sigma}_{\lambda 0}\right) $
\begin{equation}
  \bar{\mu}_{\lambda 0} = \bar{\sigma}_{\lambda 0}\left( \frac{\mu_{\lambda_0}}{\sigma^2_{\lambda_0}} + \frac{\frac{\lambda_1}{\phi_{\lambda}}}{\frac{\sigma^2_{\xi}}{\phi_{\lambda}^2}}\right)  
\end{equation}

\begin{equation}
\bar{\sigma}_{\lambda 0} = \frac{\frac{\sigma^2_{\lambda_0}\sigma^2_{\xi}}{\phi^2_{\lambda}}}{\sigma^2_{\lambda_0}  + \frac{\sigma^2_{\xi}}{\phi^2_{\lambda}}}    
\end{equation}

where $\mu_{\lambda_0}$ and $\sigma^2_{\lambda_0}$ are the prior mean and variance.

The full conditional distribution of $\boldsymbol{\lambda}$ is given by: 
      
\begin{equation}
\scriptsize
   p(\boldsymbol{\lambda} | .)  = \prod_{t=1}^T p(\lambda_t | \lambda_{t-1}, \lambda_{t+1},.) \\
\end{equation}

\begin{equation}
\scriptsize
 p(\lambda_t|.)   \propto \omega_t^{-1}(1-\delta_t^2)^{-0,5} exp \left[-\frac{1}{2} \left(\frac{y_{t} - \boldsymbol{x_{t}} \pi - \sqrt{h_{t}}\zeta_{t}  - \sqrt{h_{t} }o_{t} ^{-0.5}\omega_{t}  \delta_{t}  v_t}{\sqrt{h_{t} }o_{t} ^{-0.5} \omega_{t}  (1-\delta_{t} ^2)^{0.5}}\right)^2 - \frac{1}{2} \frac{(\lambda_t - \mu_{\lambda_t})^2 }{\sigma^2_{\lambda_t}}   \right]
\end{equation}

\begin{equation}
\begin{array}{l}\label{propnl}
 \mu_{\lambda_t} = \frac{\phi_{\lambda}}{\phi_{\lambda}^2+1} \left( \lambda_{t+1} + \lambda_{t-1} \right) \\
 \sigma_{\lambda}^2 = \frac{\sigma^2_{\xi}}{\phi_{\lambda}^2+1}
\end{array}
\end{equation}

\subsubsection{Univariate time varying skewness stochastic volatility model: skew-t shocks}

The full conditional distribution of $\{ v_{t}\}_{t=1}^T$ is given by: 
        
\begin{equation}\scriptsize
 p(v_t| .)   \propto exp \left[ -\frac{1}{2} \left( \frac{1}{(1-\delta_t^2)}v_t^2 - \frac{2o_t^{0.5}\delta_t h_t^{-0.5}}{\omega_t(1-\delta_t^2)}(y_t - \boldsymbol{x_{t}\pi} - \zeta_t\sqrt{h_t})v_t\right) \right] \mathbb{I}( 0  \leq v_t < \infty)
\end{equation}
this is a truncated normal $N \left( \frac{\delta_t o_t^{0,5}h_t^{-0.5}[y_t - \boldsymbol{x_{t}\pi}] - \delta_t o_t^{0.5}\zeta_t}{\omega_t}, 1-\delta_t^2\right)_{[0,\infty)}$

The full conditional distribution of  $\{ o_{t}\}_{t=1}^T$ is given by:  
\begin{equation} 
\scriptsize
p(o_t |.) \propto o_t^{\frac{\nu + 1}{2} - 1} exp \left[ -\frac{o_t}{2} \left( \nu_t + \frac{h_t^{-1}(y_t - \boldsymbol{x_{t}}\boldsymbol{\pi} - \sqrt{h_t}\zeta_t)^2}{ \omega_t^2(1-\delta_t^2)}  \right) \right] exp \left[  \frac{ (y_t - \boldsymbol{x_{t}}\boldsymbol{\pi} - \sqrt{h_t}\zeta_t)(h_t^{-0.5}o_t^{0.5}\delta_t v_t)}{\omega_t(1-\delta_t^2)} \right]   
\end{equation}

The full conditional distribution of $\boldsymbol{\pi}$ is Normal: 
\begin{equation}
 f(\boldsymbol{\pi}| .) \sim N (\overline{\mu}_{\pi} ,  \overline{\Sigma}_{\pi} )  \\
\end{equation}
%
%\begin{equation}
%\scriptsize
%f(\beta| .) &\propto exp \left[ - \frac{1}{2}  \sum_{t=1}^T  \frac{(\tilde{y}_t - x_{t}\beta)^2}{\sigma^2_t} \right] |\Sigma_{\beta}|^{-1} exp \left[ -\frac{1}{2} (\beta - \underline{\mu_{\beta}})'\underline{\Sigma_{\beta}^{-1}}(\beta - \underline{\mu_{\beta}}) \right]  \\
%\end{equation}
%
\begin{equation*}
\begin{array}{l}
 \overline{\mu}_{\pi} =  \overline{\Sigma}_{\pi} \left(\sum_{t=1}^T \frac{1}{\sigma^2_t} \boldsymbol{x_{t}}'\tilde{y}_t  + \underline{\Sigma_{\pi}}^{-1} \underline{\mu_{\pi}}\right)  \\
 \overline{\Sigma}_{\pi}^{-1} = \underline{\Sigma_{\pi}}^{-1} + \sum_{t=1}^T \frac{1}{\sigma^2_t} \boldsymbol{x_{t}}'\boldsymbol{x_{t}}   \\
\end{array}
\end{equation*}

where:

$ \tilde{y}_t \equiv y_t - \sqrt{h_t}\zeta_t - \sqrt{h_t}o_t^{-0.5}\omega_t \delta_t v_t$

$  \sigma^2_t \equiv h_t \omega^2_to_t^{-1}(1-\delta_t^2)$

The full conditional distribution of $h_t$ is given by: 

\begin{equation}
\scriptsize
p(\boldsymbol{h} | .)  = \prod_{t=1}^T p(h_t | h_{t-1}, h_{t+1},.)     
\end{equation}

\begin{equation}
\scriptsize
p(h_t|.)  \propto h_t ^{-1,5}  exp \left[\left(-\frac{1}{2}\left(\frac{y_{t} - \boldsymbol{x_{t}} \boldsymbol{\pi} - \sqrt{h_{t}}\zeta_{t}  - \sqrt{h_{t} }o_{t} ^{-0.5}\omega_{t}  \delta_{t}  v_{t}}{\sqrt{h_{t} }o_{t} ^{-0.5} \omega_{t}  (1-\delta_{t} ^2)^{0.5}}\right)^2  -\frac{1}{2} \frac{(ln h_t - \mu_{h_t})^2}{\sigma_{h_t}^2}       \right)  \right]
\end{equation}

\begin{equation}\label{prop}
\begin{aligned}{l}
& \mu_{h_t} = \frac{\phi_{h}}{\phi_{h}^2 + 1} \left( ln h_{t+1} + ln h_{t-1} \right) \\
& \sigma_{h}^2 = \frac{\sigma^2_{\eta}}{\phi_{h}^2 + 1}
\end{aligned}
\end{equation}

The full conditional distribution of $\lambda_t$ is given by: 
      
\begin{equation}
\scriptsize
     p(\boldsymbol{\lambda} | .)  = \prod_{t=1}^T p(\lambda_t | \lambda_{t-1}, \lambda_{t+1},.) 
\end{equation}
\begin{equation}
\scriptsize
p(\lambda_t|.)  \propto \omega_t^{-1}(1-\delta_t^2)^{-0,5} exp \left[-\frac{1}{2} \left(\frac{y_{t} - \boldsymbol{x_{t}} \pi - \sqrt{h_{t}}\zeta_{t}  - \sqrt{h_{t} }o_{t} ^{-0.5}\omega_{t}  \delta_{t}  v_t}{\sqrt{h_{t} }o_{t} ^{-0.5} \omega_{t}  (1-\delta_{t} ^2)^{0.5}}\right)^2 - \frac{1}{2} \frac{(\lambda_t - \mu_{\lambda_t})^2 }{\sigma^2_{\lambda_t}}   \right]    
\end{equation}
\begin{equation}\label{propl}
\begin{aligned}{l}
\scriptsize
& \mu_{\lambda_t} = \frac{\phi_{\lambda}}{\phi_{\lambda}^2+1} \left(\lambda_{t+1} + \lambda_{t-1} \right) \\
& \sigma_{\lambda}^2 = \frac{\sigma^2_{\xi}}{\phi_{\lambda}^2+1}
\end{aligned}
\end{equation}

The full conditional distribution of $\boldsymbol{v}={v_1, \ldots,v_T}$ is given by: 
        
\begin{equation}
 \scriptsize
  p(\boldsymbol{v} | .)  = \prod_{t=1}^T p(v_t |.)  
\end{equation}

\begin{equation}
\scriptsize
 p(v_t| .)   \propto exp \left[ -\frac{1}{2} \left( \frac{1}{(1-\delta_t^2)}v_t^2 - \frac{2o_t^{0.5}\delta_t h_t^{-0.5}}{\omega_t(1-\delta_t^2)}(y_t - \boldsymbol{x_{t}}\boldsymbol{\pi} - \zeta_t\sqrt{h_t})v_t\right) \right] \mathbb{I}( 0  \leq v_t < \infty)   
\end{equation}

this is is a \textit{Truncated Normal} $ \left( \frac{\delta_t o_t^{0,5}h_t^{-0.5}[y_t - \boldsymbol{x_{t}}\boldsymbol{\pi}] - \delta_t o_t^{0,5}\zeta_t}{\omega_t}, 1-\delta_t^2\right)_{[0,\infty)}$

%\begin{equation}
%    q(o_t) = \frac{\left( \frac{1}{2} \left[ \nu_t + \frac{h_t^{-1}(y_t - x_{t}\beta - \sqrt{h_t}\zeta)^2}{\omega_t^2(1-\delta_t^2)} \right]\right)^{\frac{\nu +1}{2}} o_t^{\frac{\nu +1}{2} - 1 }}{\Gamma(\frac{\nu_t +1}{2})} exp \left[ -o_t\frac{1}{2} \left( \nu + \frac{h_t^{-1}(y_t - x_{t}\beta - \sqrt{h_t}\zeta_t)^2}{ \omega_t^2(1-\delta_t^2)}  \right) \right]
%\end{equation}

\subsubsection{VAR with Skew Normal shocks}

The full conditional distribution of $vec(\boldsymbol{\Pi})$ is $\mathcal{N}( vec(\boldsymbol{\bar{\mu}_{\Pi}}), \boldsymbol{\bar{V}_{\Pi}} )$, where:
\begin{equation}
\scriptsize
\boldsymbol{\bar{\mu}_{\Pi}} =  \boldsymbol{\bar{V}_{\Pi}} \left[ vec \left( \sum_{t=1}^T\boldsymbol{X_t\tilde{y}_t'\Sigma_t^{-1}} \right) + \boldsymbol{\uline{V_{\Pi}}}^{-1}vec( \boldsymbol{\uline{\mu_{\Pi}}})  \right]
\end{equation}

with $\boldsymbol{\tilde{y}}_t \equiv \boldsymbol{y}_t - \boldsymbol{H_t^{0.5}}\boldsymbol{A^{-1}}\boldsymbol{\zeta}_t  - \boldsymbol{H_t^{0.5}}\boldsymbol{A^{-1}}\boldsymbol{\Omega_t}\boldsymbol{\Delta_t}\boldsymbol{v_t}$ and: 
\begin{equation}
\scriptsize
\boldsymbol{\bar{V}_{\Pi}} =   \boldsymbol{\uline{V_{\Pi}}}^{-1} + \sum_{t=1}^T( \boldsymbol{\Sigma_t^{-1}} \otimes \boldsymbol{X_tX_t'})
\end{equation}
where $\boldsymbol{\Sigma}_t \equiv \boldsymbol{A}^{-1} \boldsymbol{H}_t \boldsymbol{\omega_t}^2 (\boldsymbol{I} - \boldsymbol{\Delta_t}^2)\boldsymbol{A'}^{-1}$ while $\boldsymbol{\uline{\mu_{\Pi}}}$ and $\boldsymbol{\uline{V_{\Pi}}}$ are the prior mean and variance covariance matrix.

The full conditional distribution of the elements in $\boldsymbol{A}$ is derived adapting our framework to the approach of \citet{COGLEY2005262}. Considering the system: 
\begin{equation}
\boldsymbol{Au_t} = \boldsymbol{H_t^{0.5}\varepsilon_t}
\end{equation}
%
%\begin{bmatrix}
%1 & 0 &\ldots &   0             \\
%a_{21} & 1 & \ddots & 0         \\
%\vdots & \ddots & \ddots & 0    \\
%a_{n1} & \ldots & a_{n, n-1} & 1\\
%\end{bmatrix}
%\begin{bmatrix}
%u_{1t}      \\
%\vdots      \\
%\vdots      \\
%u_{nt} \\
%\end{bmatrix}
%\]
%
since $\varepsilon_{it} = \zeta_{it} + \omega_{it}\delta_{it}v_{it} + \omega_{it}\sqrt{1 - \delta_{it}^2}z_{it}$ we have : 

\begin{equation} \label{system}
\scriptsize
\begin{array}{l}
u_{1t} = \sqrt{h}_{1t}(\zeta_{1t} +  \omega_{1t}\delta_{1t}v_{1t} + \omega_{1t}\sqrt{1 - \delta_{1t}^2}z_{1t}) \\
u_{2t} = - a_{21}u_{1t} + \sqrt{h}_{2t}(\zeta_{2t} +  \omega_{2t}\delta_{2t}v_{2t} + \omega_{2t}\sqrt{1 - \delta_{2t}^2}z_{2t}) \\
u_{3t}  = -a_{31}u_{1t}  -a_{32}u_{2t} + \sqrt{h}_{3t}(\zeta_{3t} +  \omega_{3t}\delta_{3t}v_{3t} + \omega_{3t}\sqrt{1 - \delta_{3t}^2}z_{3t}) \\ 
\hspace{0.25cm} \vdots \hspace{2cm} \vdots \hspace{2cm} \vdots \\
u_{Nt}  = -a_{N1}u_{1t}  -a_{N2}u_{2t} \ldots  -a_{N,N-1}u_{2t}  + \sqrt{h}_{Nt}(\zeta_{Nt} +  \omega_{Nt}\delta_{Nt}v_{Nt} + \omega_{Nt}\sqrt{1 - \delta_{Nt}^2}z_{Nt}) 
\end{array}
\end{equation}
therefore :
\begin{equation*}
\scriptsize
\begin{array}{l}
u_{1t} - \sqrt{h}_{1t}(\zeta_{1t} +  \omega_{1t}\delta_{1t}v_{1t}) = \sqrt{h}_{1t}\omega_{1t}\sqrt{1 - \delta_{1t}^2}z_{1t} \\
u_{2t} - \sqrt{h}_{2t}(\zeta_{2t} +  \omega_{2t}\delta_{2t}v_{2t}) = - a_{21}u_{1t}  + \sqrt{h}_{2t}\omega_{2t}\sqrt{1 - \delta_{2t}^2}z_{2t} \\
u_{3t} -\sqrt{h}_{3t}(\zeta_{3t} +  \omega_{3t}\delta_{3t}v_{3t}) = -a_{31}u_{1t}  - a_{32}u_{2t}  + \sqrt{h}_{3t}\omega_{3t}\sqrt{1 - \delta_{3t}^2}z_{3t} \\
\hspace{0.25cm} \vdots \hspace{2cm} \vdots \hspace{2cm} \vdots \\
u_{Nt} - \sqrt{h}_{Nt}(\zeta_{Nt} +  \omega_{Nt}\delta_{Nt}v_{Nt}) = -a_{N1}u_{1t}  -a_{N2}u_{2t} \ldots  -a_{N,N-1}u_{2t}   + \sqrt{h}_{Nt}\omega_{Nt}\sqrt{1 - \delta_{Nt}^2}z_{Nt}
\end{array}
\end{equation*}

Since I condition on the parameters, the mixing variables and the latent states I can define $\tilde{u}_{it} = u_{it} - \sqrt{h}_{it}(\zeta_{it} +  \omega_{it}\delta_{it}v_{it})$ for $i = 1, \ldots, N$  and $\tilde{\sigma}^2_{it} = \sqrt{h_{it}}\omega_{it}\sqrt{1- \delta_{it}^2}$ and derive the full conditional posterior for the elements of $\boldsymbol{A}$ by exploiting the system of equations: 

\begin{equation}\label{sistem}
\scriptsize
\begin{array}{l}
\tilde{u}_{1t}  = \tilde{\sigma}^2_{1t}z_{1t} \\
\tilde{u}_{2t}  = - a_{21}u_{1t}  + \tilde{\sigma}^2_{2t}z_{2t} \\
\tilde{u}_{3t}  = -a_{31}u_{1t}  - a_{32}u_{2t}  + \tilde{\sigma}^2_{3t}z_{3t} \\
\hspace{0.25cm} \vdots \hspace{2cm} \vdots \hspace{2cm} \vdots \\
\tilde{u}_{Nt}  = -a_{N1}u_{1t}  -a_{N2}u_{2t} \ldots  -a_{N,N-1}u_{2t}   + \tilde{\sigma}^2_{Nt}z_{Nt}
\end{array}
\end{equation}

where $z_{it} \sim \mathcal{N}(0,1)$. Assuming a \textit{Normal} prior for the elements in $\boldsymbol{A}$ and defining $\boldsymbol{a}_i$ the vector that collects the free elements in the $i^{th}$ row of the $\boldsymbol{A}$ matrix, I can use standard linear regression results to show that the full conditional posterior of $\boldsymbol{a_i}$ is given by $\boldsymbol{a_i}\sim\mathcal{N}(\boldsymbol{\bar{\mu}_{a,i}} ,\boldsymbol{\bar{V}_{a,i}})$ where: 

\begin{equation}
 \begin{array}{l}
\boldsymbol{\bar{\mu}_{a,i}} = \boldsymbol{\bar{V}_{a,i}}(\boldsymbol{\uline{V_{a,i}}}^{-1} \boldsymbol{\uline{\mu_a}} + \sum_{t=1}^T \tilde{\sigma_{it}}^{2^-1}  \boldsymbol{u_{it}}'\tilde{u}_{it}    )\\
\boldsymbol{\bar{V}_{a,i}} = (\boldsymbol{\uline{V_{a,i}}}^{-1}  + \sum_{t=1}^T \tilde{\sigma_{it}}^{2^{-1}}\boldsymbol{u_{it}}'\boldsymbol{u_{it}})^{-1}
 \end{array}
\end{equation}

where $\boldsymbol{u_{it}}$ is the vector colleting the right hand variables of the $i^{th}$ equation in the system above (\ref{sistem}) with $i = 2, \ldots, N$ and $\boldsymbol{\uline{\mu_a}}$ and $\boldsymbol{\uline{V_{a,i}}}$ are the prior mean and variance covariance matrix of the free elements of the $i^{th}$ row of $\boldsymbol{A}$.

\subsubsection{VAR with Skew-t shocks}

The full conditional distribution of $vec(\boldsymbol{\Pi})$ is $\mathcal{N}( vec(\boldsymbol{\bar{\mu}_{\Pi}}), \boldsymbol{\bar{V}_{\Pi}} )$, where:
\begin{equation}
\scriptsize
\boldsymbol{\bar{\mu}_{\Pi}} =  \boldsymbol{\bar{V}_{\Pi}} \left[ vec \left( \sum_{t=1}^T\boldsymbol{X_t\tilde{y}_t'\Sigma_t^{-1}} \right) + \boldsymbol{\uline{V_{\Pi}}}^{-1}vec( \boldsymbol{\uline{\mu_{\Pi}}})  \right]
\end{equation}
with $\boldsymbol{\tilde{y}}_t \equiv \boldsymbol{y}_t - \boldsymbol{H_t^{0.5}}\boldsymbol{A^{-1}}\boldsymbol{\zeta}_t  - \boldsymbol{H_t^{0.5}}\boldsymbol{A^{-1}}\boldsymbol{\Omega_t} \boldsymbol{\Delta_t}\boldsymbol{O_t}^{-0.5} \boldsymbol{v_t}$
and: 
\begin{equation}
\scriptsize
\boldsymbol{\bar{V}_{\Pi}} =   \boldsymbol{\uline{V_{\Pi}}}^{-1} + \sum_{t=1}^T( \boldsymbol{\Sigma_t^{-1}} \otimes \boldsymbol{X_tX_t'})
\end{equation}
where $\boldsymbol{\Sigma_t} \equiv \boldsymbol{A^{-1}}\boldsymbol{H}_t \boldsymbol{\omega_t^2}(\boldsymbol{I} - \boldsymbol{\Delta_t}^2)\boldsymbol{O_t}^{-1}\boldsymbol{A'^{-1}}$ while $\boldsymbol{\uline{\mu_{\Pi}}}$ and $\boldsymbol{\uline{V_{\Pi}}}$ are the prior mean and variance covariance matrix.  \\

The full conditional for $\boldsymbol{A}$ is derived following the same steps in the VAR with  \textit{Skew-normal} shocks just by considering that (\ref{system}) becomes: 
\begin{equation} 
\scriptsize
\begin{array}{l}
u_{1t} = \sqrt{h}_{1t}(\zeta_{1t} +  \omega_{1t}\delta_{1t}o^{-0.5}_{1t}v_{1t} + \omega_{1t}\sqrt{1 - \delta_{1t}^2}o^{-0.5}_{1t}z_{1t}) \\
u_{2t} = - a_{21}u_{1t} + \sqrt{h}_{2t}(\zeta_{2t} +  \omega_{2t}\delta_{2t}o^{-0.5}_{2t}v_{2t} + \omega_{2t}\sqrt{1 - \delta_{2t}^2o^{-0.5}_{2t}}z_{2t}) \\
u_{3t}  = -a_{31}u_{1t}  -a_{32}u_{2t} + \sqrt{h}_{3t}(\zeta_{3t} +  \omega_{3t}\delta_{3t}o^{-0.5}_{3t}v_{3t} + \omega_{3t}\sqrt{1 - \delta_{3t}^2}o^{-0.5}_{3t}z_{3t}) \\ 
\hspace{0.25cm} \vdots \hspace{2cm} \vdots \hspace{2cm} \vdots \\
u_{Nt}  = -a_{N1}u_{1t}  -a_{N2}u_{2t} \ldots  -a_{N,N-1}u_{2t}  + \sqrt{h}_{Nt}(\zeta_{Nt} +  \omega_{Nt}\delta_{Nt}o^{-0.5}_{Nt}v_{Nt} + \omega_{Nt}\sqrt{1 - \delta_{Nt}^2}o^{-0.5}_{Nt}z_{Nt}) 
\end{array}
\end{equation}

\subsection{Metropolis Hastings Step to draw the mixing variable \textit{o}} \label{draw}
In the time varying skewness stochastic volatility models with \textit{Skew-t} shocks, the full conditional distribution of  $\{ o_{t}\}_{t=1}^T$ is given by:  
\begin{equation} 
\scriptsize
p(o_t |.) \propto o_t^{\frac{\nu + 1}{2} - 1} exp \left[ -\frac{o_t}{2} \left( \nu_t + \frac{h_t^{-1}(y_t - \boldsymbol{x_{t}}\boldsymbol{\pi} - \sqrt{h_t}\zeta_t)^2}{ \omega_t^2(1-\delta_t^2)}  \right) \right] exp \left[  \frac{ (y_t - \boldsymbol{x_{t}}\boldsymbol{\pi} - \sqrt{h_t}\zeta_t)(h_t^{-0.5}o_t^{0.5}\delta_t v_t)}{\omega_t(1-\delta_t^2)} \right]   
\end{equation}

Since it is not possible to directly sample from this full conditional distribution, I use Metropolis Hastingss step to draw from this conditional distribution. I use as proposal

$Gamma \left( \frac{\nu +1}{2}, \frac{1}{2} \left[ \nu + \frac{h_t^{-1}(y_t - \boldsymbol{x_{t}}\boldsymbol{\pi} - \sqrt{h_t}\zeta)^2}{\omega_t^2(1-\delta_t^2)} \right]\right)$.
The acceptance probability in the Metropolis Hastings step is:
\begin{equation}
p =  exp \left[\frac{(y_t - \boldsymbol{x_{t}\pi} - \sqrt{h_t}\zeta_t)h_t^{-0.5}o_t^{*^{0.5}}\delta_t v_t}{\omega_t(1-\delta_t^2)} - \frac{ (y_t - \boldsymbol{x_{t}\pi} - \sqrt{h_t}\zeta_t)h_t^{-0.5}o^{m_t^{0.5}}\delta_t v_t}{\omega_t(1-\delta_t^2)}\right]\\
\end{equation}

where $o^*$ is a new draw from the proposal and $o^m$ is the previous draw.

\subsection{Particle Step in the Gibbs Sampler}\label{APSTEP}

Table \ref{tab:part} presents the details on the \textit{Particle Step} used in the \textit{Gibbs Sampler} to draw the volatilities and the shape parameters. $s_t$ stands for the generic unobserved latent state being $log(h_t)$ in Step 5) and $\lambda_t$ in Step 7) of the Gibbs Sampler in Table \ref{tab:Gibbs sampler}. As anticipated above a valid particle approximation to the Gibbs Sampler requires a Conditional Sequential Monte Carlo update which guarantees that a pre-specified path of the state variables is ensured to survive all the resampling steps \citep{andrieu2010particle}. Hence, if I consider a generic iteration $ m + 1 $ of the Gibbs Sampler, when using $K$ particles to approximate $p(h_{1}, \ldots, h_{T}| \boldsymbol{\Theta},\boldsymbol{v}, \boldsymbol{\lambda})$ and $p(\lambda_{i1}, \ldots, \lambda_{iT}| \boldsymbol{\Theta}, \boldsymbol{v}, \boldsymbol{h}) $ ,  only $K-1$ particles are generated while the $K^{th}$ particle is set to the pre-specified path $ h_{1:T}^{(m)}$ and $ \lambda_{1:T}^{(m)}$. In the particle approximation I use the transition equations (\ref{logv}) and (\ref{lambdat}) as importance densities $g_{\theta}(s_t)$ and compute the weights accordingly. I refer to the original paper, \citet{lindsten2014particle} for the details on the \textit{Ancestral Sampling} step, that for $t>2$ artificially assign a history to the partial pre-specified path  $s_{t:T}^{(m)}$. 

%namely: 

%\begin{equation}
%    w_{t}^k = f(y_t|s_t, .)\hspace{1cm} \text{for} \hspace{0.2cm} k = 1,\ldots, K
%\end{equation}
%
%where $ f(y_t|s_t, .)$ is given the Gaussian distribution, obtained by conditioning on the mixing variables, the parameters and the other states in the model. 

\begin{table}[H]
    \centering
    \caption{Particle Step in the Gibbs Sampler}
    \begin{tabular}{l l}
    \hline 
    \small \textit{Particle Step with Ancestor Sampling}  &  \\  
    \hline\hline
\scriptsize \hspace{1.5cm} Draw $s_{1}^k \sim g_\theta(s_{1}) $ & \scriptsize for  $ k = 1, \ldots, K-1$  \\
\scriptsize \hspace{1.5cm} Set $s_{1}^K = s_{1}^{(m)}$  \\
\scriptsize \hspace{1.5cm} Compute  $ w^k_1 = W_{1}(s_{1}^k)$ and normalize the weights & \scriptsize for $ k = 1, \ldots, K$ \\
\scriptsize \hspace{1.5cm} for $t = 2 : T$    \\
\scriptsize \hspace{2cm} Re-sampling step: sample $\{s_{t-1}^{k}\}_{k=1}^{K}$ with probabilities given by $\{w_{t-1}^{k}\}_{k=1}^{K}$    \\
\scriptsize \hspace{2cm} Draw $s_{t}^k \sim g_\theta(s_{t})$ & \scriptsize for $ k = 1, \ldots, K-1$  \\   
\scriptsize \hspace{2cm} Set $s_{t}^K = s_{t}^{(m)}$  \\
\scriptsize \hspace{2cm} Ancestral sampling step     \\ 
\scriptsize \hspace{2cm} Compute $ w^k_t = W_{t}(s_{t}^k)$ and normalize the weights& \scriptsize for $ k = 1, \ldots, K$ \\
\scriptsize \hspace{1.5cm} end    \\
\scriptsize \hspace{1.5cm} Draw $j$ with $Pr(j = k) \propto $ $w_T^k$     \\
    \hline \hline
    \end{tabular}
    \label{tab:part}
\end{table}

In alternative to the particle step, it can also be considered an independence Metropolis Hastings step to draw the log-volatilities and the shape parameters. In particular, I considered a log-normal proposal density for the volatility  (on the lines of \citet{COGLEY2005262}) as:

\begin{equation}
q(h_t) \propto h_t^{-1}exp\left[ - \frac{(lnh_t - \mu_{h_t})^2}{2\sigma_{h}^2}    \right]
\end{equation}

with $\mu_{h_t}$ and $\sigma_{h}$ defined in equations (\ref{propn})  for the \textit{Skew-Normal} case and (\ref{prop}) for the \textit{Skew-t}. The acceptance probabilities in the model with \textit{Skew-Normal}  and \textit{Skew-t} shocks  are respectively given by: %
  \begin{equation}
     p = \frac{h_t^{*^{-0,5}} exp \left[ -\frac{(y_{t} - \boldsymbol{x_t\pi} - \sqrt{h^*_{t}}\zeta_{t}  - \sqrt{h^*_{t} }\omega_{t}  \delta_{t}  v_t)^2}{2h_t^*\omega_t^2(1-\delta_t^2)}  \right]}{h^{m_t^{-0,5}} exp\left[ - \frac{(y_{t} - \boldsymbol{x_t\pi} - \sqrt{h_{t}^{m}}\zeta_{t}  - \sqrt{h_{t}^{m}}\omega_{t}  \delta_{t}  v_t)^2}{2h_t^{m}\omega_t^2(1-\delta_t^2)}  \right]}
  \end{equation}
  
  \begin{equation}
 p = \frac{h_t^{*^{-0,5}} exp \left[ -\frac{(y_{t} - \boldsymbol{x_t\pi} - \sqrt{h^*_{t}}\zeta_{t}  - \sqrt{h_{t}^* }o_t ^{-0.5}\omega_{t}  \delta_{t}  v_t)^2}{2h^*_to_t^{-1}\omega_t^2(1-\delta_t^2)}  \right]}{h^{m_t^{-0,5}} exp\left[ - \frac{(y_{t} - \boldsymbol{x_t\pi} - \sqrt{h_{t}^{m}}\zeta_{t}  - \sqrt{h_{t}^{m}}o_t ^{-0.5}\omega_{t}  \delta_{t}  v_t)^2}{2h_t^{m}o_t^{-1}\omega_t^2(1-\delta_t^2)}  \right]}
 \end{equation}
where $h_t^*$ is the new draw from the proposal distribution, while $h_t^m$ is the previous draw.
Instead, for the shape parameters, I considered a Normal proposal: 
\begin{equation}
 q(\lambda_t) \sim N (\mu_{\lambda_t}, \sigma^2_{\lambda}) 
\end{equation}
with $\mu_{h_t}$ and $\sigma_{h}$ defined in equations (\ref{propnl})  for the \textit{Skew-Normal} case and (\ref{propl}) for the \textit{Skew-t}.
The acceptance probabilities in the model with \textit{Skew-Normal}  and \textit{Skew-t} shocks are respectively given by: 
\begin{equation}
p = \frac{\omega_t^{*^{-1}}(1-\delta_t^{*^2})^{-0,5} exp \left[-\frac{(y_{t} - \boldsymbol{x_t\pi} - \sqrt{h_{t}}\zeta_{t}^*  - \sqrt{h_{t} }\omega^*_{t}  \delta^*_{t}  v_t)^2}{2h_t\omega_t^{*^2}(1-\delta_t^{*^2})} \right]}{(\omega_t^m)^{-1}(1-\delta_t^{m^2})^{-0,5} exp \left[-\frac{(y_{t} - \boldsymbol{x_t\pi} - \sqrt{h_{t}}\zeta^m_{t}  - \sqrt{h_{t}}\omega^m_{t}  \delta^m_{t}  v_t)^2}{2h_t\omega_t^{m^2}(1-\delta_t^{m^2})} \right]}
\end{equation}

\begin{equation}
p = \frac{\omega_t^{*^{-1}}(1-\delta_t^{*^2})^{-0,5} exp \left[-\frac{(y_{t} - \boldsymbol{x_t\pi} - \sqrt{h_{t}}\zeta_{t}^*  - \sqrt{h_{t} }o_t^{-0.5}\omega_{t}^*  \delta^*_{t}  v_t)^2}{2h_to_t^{-1}\omega_t^{*^2}(1-\delta_t^{*^2})} \right]}{(\omega_t^m)^{-1}(1-\delta_t^{m^2})^{-0,5} exp \left[-\frac{(y_{t} - \boldsymbol{x_t\pi} - \sqrt{h_{t}}\zeta^m_{t}  - \sqrt{h_{t}}o_t ^{-0.5}\omega^m_{t}  \delta^m_{t}  v_t)^2}{2h_to_t^{-1}\omega_t^{m^2}(1-\delta_t^{m^2})} \right]}
\end{equation}
where $\omega_t^*, \zeta_t^*, \delta_t^*$ are functions of the new draw from the proposal  $\lambda_t^*$, while $\omega_t^m, \zeta_t^m, \delta_t^m$ are $\lambda_t^m$ are functions of the previous draw $\lambda_t^m$.

\section{Appendix}

\subsection{Priors and hyper-parameters}\label{sec:priors}
Table \ref{priors} and Table \ref{priorsvar} report the specification of the priors and the choice of the hyper-parameters used for the estimation of  the models in the empirical application.

\begin{table}[H]
    \centering    \caption{Priors for the parameters of the TVSSV model}
    \begin{tabular}{c|cc|}
    \hline
        Parameter & Prior  \\ \hline \hline
        $\sigma^2_{\xi}$ &  \textit{InverseGamma}$\left(5,0.16\right)$\\
        $\sigma^2_{\eta}$ & \textit{InverseGamma}$\left(5,0.16\right)$\\
        $\phi_{h,\lambda}$ & $\mathcal{N}(1, 0.01)$ \\
        $\beta_{1}$ & $\mathcal{N}(0, 10)$ \\
        $\pi_i$ & $\mathcal{N}(\underline{\mu_{\pi}}, \underline{\sigma_{\pi,i}})$ \\      
        $log(h_0)$ & $\mathcal{N}\left(\hat{h}_0,100\right)$\\
        $\lambda_0$ & $\mathcal{N}\left(0,10\right)$\\
    \hline \hline
    \end{tabular}
    \label{priors}
\end{table}\noindent
$\hat{h}_{i,0}$ is the estimated variance from an AR(4) model to each series using an initial sample of 40 observations. In the application in Section \ref{sec:gar} I assume that the elements of $\boldsymbol{\pi}$ namely $\pi_{i}$ are centered in zero, namely $\mu_{\pi,i} = 0$ and the variances $\sigma_{\pi,i}$ are set following \citet{ccm2015}.
 %\citep{L1986} such that the prior for the variance of the coefficient of the generic $l$ lag of the $j^{th}$ regressor in the $i^{th}$  equation is : 
% 
For the VAR I consider the following priors: 
\begin{table}[H]
    \centering    \caption{Priors for the parameters of the VAR TVSSV model}\label{priorsvar}
    \begin{tabular}{c|cc|}
    \hline
        Parameter & Prior  \\ \hline \hline
        $vec(\boldsymbol{\Pi})$ & $\mathcal{N}(vec(\boldsymbol{\uline{\mu_{\Pi}}}), \boldsymbol{\uline{V_{\Pi}}})$\\ 
        $a_{ij}$ & $\mathcal{N}(0,100)$\\
    \hline \hline
    \end{tabular}
\end{table}
 
where the elements of $vec(\boldsymbol{\uline{\mu_{\Pi}}})$ are equal to zero for the coefficients on the cross-equation lags and for the intercept. The coefficients of the own lags are centered in 0  for stationary variables and on 1 for non-stationary variables.

$\uline{V_{\Pi}}$ has the Minnesota type prior: 
\begin{equation}
  v_{ij,l} = 
\begin{cases*}
 & $\frac{\theta_1}{l^{\theta_4}}$ \hspace{2.5cm} if $ \quad i = j$  \\
 & $\frac{\sigma^2_i\theta_1\theta_2}{\sigma^2_{j}l^{\theta_4}}$ \hspace{2cm} if $ \quad i \neq j$ \\
\end{cases*}    
\end{equation}

 where I set $\theta_1 = 0.04 \quad \theta_2 = 0.025 \quad \theta_3 = 100 \quad \theta_4 = 2 $. We estimate $\sigma^2_i$ from univariate AR(12) regressions. 

\subsection{Variables in the medium scale VAR}
\begin{table}[htp!]
\caption{Variable transformations}
\begin{tabular}{l|c}
{\color[HTML]{000000} \textbf{Variable}}                 & \textbf{Transformation} \\
\hline \hline
Real personal consumption expenditures                   & $ log  $         \\
Industrial Production                           & $ log  $         \\
Unemployment Rate                               & $ level$               \\
Avg Weekly Hours Worked                                  & $log$                     \\
Consumer Price Index                                     & $ log  $         \\
Fed Funds Rate                                           & $ level$               \\
10-Year Treasury Yield -  Fed Funds Rate                 & $ level $                  \\
Moody’s Baa Corporate Bond Yield - the   Fed Funds Rate  & $ level $               \\
Standard and Poors index                                 & $ log  $         \\
\hline \hline
\end{tabular}
\label{tab:trans}
\end{table}

\subsection{Other Figures}\label{otherf}

\begin{figure}[H]
    \caption{CRPS and Tail Weighted CRPS (left tail)}   \centering
    \includegraphics[scale = 0.5]{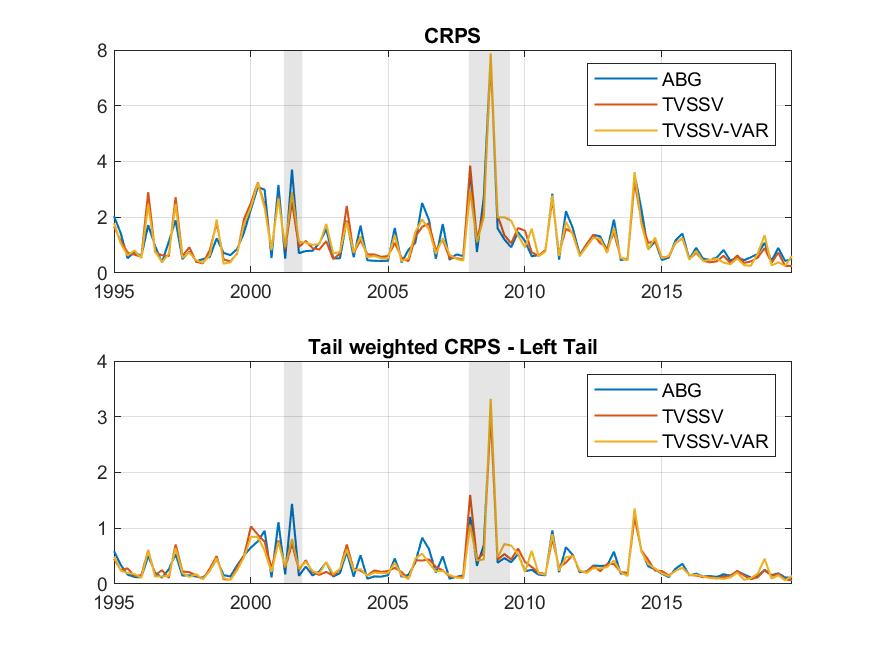}
    \label{fig:crps_TS}
    \floatfoot{\scriptsize Note:  The figure above shows the time series of the Cumulative Ranked Probability Scores (CRPS), while the figure below shows the time series of the Left Tail Weighted CRPS \citep{TWCRS}. In blue estimates from the two step quantile regression based method by \citep{ABG}, in red from the TVSSV univariate model with \textit{Skew-t} shocks and  in yellow the estimates from the TVSSV VAR model. }
\end{figure}

\clearpage

\begin{landscape}
\begin{center}
\begin{figure}[H]
    \centering
    \caption{Probability Integral Transforms}
    \includegraphics[scale = 0.5]{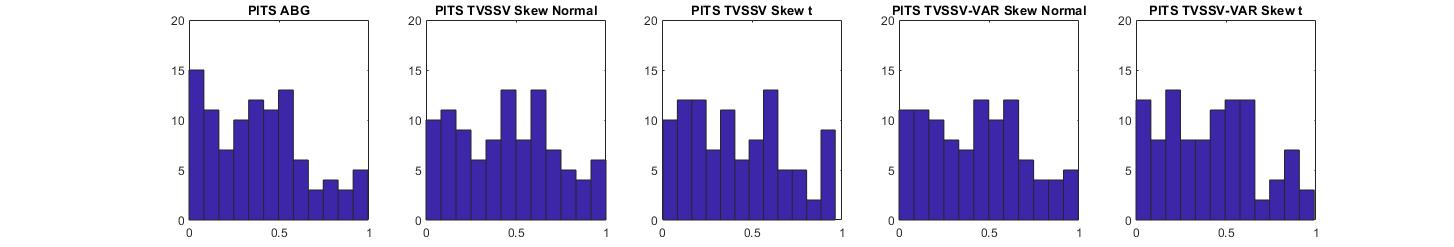}
    \label{fig:PITs}
    \floatfoot{\scriptsize Note:  Probability Integral Transforms of the forecasts from the quantile regression based method, the univariate time varying skewness stochastic volatility models with \textit{Skew-Normal} and \textit{Skew-t} shocks, and the VARs with varying skewness and stochastic volatility with \textit{Skew-Normal} and \textit{Skew-t} shocks }
\end{figure}    
\end{center}
\end{landscape}

\clearpage

\end{document}